\documentclass[a4paper,11pt]{article}
\pdfoutput=1 

\usepackage{jheppub} 
\usepackage{empheq}
\usepackage{bm}
\usepackage{slashed}
\usepackage[T1]{fontenc} 
\usepackage[all]{xy}
\usepackage{rotating}
\usepackage{amsmath} 
\usepackage{mathtools}
\usepackage{xcolor}
\usepackage{ulem}

\DeclareMathOperator{\Tr}{Tr}

\usepackage{slashed}

\def\CA{{\cal A}}

\def\CO{{\cal O}}

\def\beq#1\eeq{\begin{align}#1\end{align}}

\newcommand{\be}{\begin{equation}}
\newcommand{\ee}{\end{equation}}
\def\bea{\begin{eqnarray}}
\def\eea{\end{eqnarray}}

\preprint{LCTP-19-09}

\title{\bf \Large Precision Microstate Counting for the Entropy of Wrapped M5-branes}

\author[a]{Dongmin Gang,}
\author[b,c]{Nakwoo Kim,}
\author[d,e]{and Leopoldo A. Pando Zayas}

\affiliation[a]{Quantum Universe Center, Korea Institute for Advanced Study, Seoul 02455, Korea}
\affiliation[b]{Department of Physics and Research Institute of Basic Science, Kyung Hee University, Seoul 02447, Republic of Korea}
\affiliation[c]{School of Physics, Korea Institute for Advanced Study, Seoul 02445, Republic of Korea}
\affiliation[d]{Leinweber Center for Theoretical Physics, Randall Laboratory of Physics
\\
The University of Michigan, Ann Arbor, MI 48109-1040}
\affiliation[e]{The Abdus Salam International Centre for Theoretical Physics,  Strada Costiera 11, 34014 Trieste, Italy}

\abstract{We study  the large   $N$ expansion of  twisted partition functions of 3d $\mathcal{N}=2$ superconformal field theories arising from  $N$ M5-branes wrapped on a hyperbolic 3-manifold, $M_3$. Via the  3d-3d correspondence, the partition functions  of these 3d ${\cal N}=2$ superconformal field theories are related to simple topological invariants on the 3-manifold. The partition functions can be expressed using only  classical and one-loop perturbative invariants of  $PSL(N,\mathbb{C})$  Chern-Simons theory around irreducible flat connections on $M_3$. Using  mathematical results on the asymptotics  of the  invariants, we compute the twisted partition functions in the large $N$ limit including {\it perturbative corrections to all orders in $1/N$}. Surprisingly, the perturbative expansion terminates at finite order.  
The leading part of the partition function is of order $N^3$ and  agrees with the Bekenstein-Hawking entropy  of the dual black holes. 
The subleading part, in particular the $\log N$-terms in the field theory partition function is found to precisely match the one-loop quantum corrections in the dual eleven dimensional supergravity. The field theory results of other terms in $1/N$ provide a stringent prediction for higher order corrections in the holographic dual, which is M-theory. }

\begin{document} 
\maketitle
\flushbottom

\section{Introduction}

The mathematical equivalence of a field theory to a theory containing gravity in the context of the AdS/CFT correspondence  \cite{Maldacena:1997re}  has always been viewed as potentially a direct way to uncover intricate and intuition-defying aspects of gravity. First among these are issues related to black hole physics and, in particular, the microscopic understanding of black hole entropy. 

Recently a remarkable result has been obtained providing a microscopic understanding of the entropy of certain magnetically charged,  asymptotically $AdS_4$ black holes in the context of  AdS$_4$/CFT$_3$ \cite{Benini:2015eyy}. This impressive achievement has been extended to various situations including dyonic black holes   \cite{Benini:2016rke}, black holes with hyperbolic horizons  \cite{Cabo-Bizet:2017jsl}, black holes in massive IIA theory    \cite{Benini:2017oxt,Hosseini:2017fjo}  and to certain black holes in universal sectors of  higher-dimensional embeddings \cite{Azzurli:2017kxo}; for a review and a complete list of references, see \cite{Zaffaroni:2019dhb}.

Most of these results have been propelled by an improved understanding of three-dimensional ${\cal N}=2$ supersymmetric field theories thanks to supersymmetric localization, see \cite{Kapustin:2009kz,Jafferis:2010un,Hama:2010av} for the original developments and  \cite{Benini:2015noa,Benini:2016hjo,Closset:2016arn,Closset:2017zgf} for some recent relevant applications. The stringy origin of many of those field theories can be tracked to M2 and D2 brane configurations.  For example, the marquee case worked out in detail in  \cite{Benini:2015eyy} exploits the duality between a Chern-Simons matter theory known as ABJM  \cite{Aharony:2008ug}  and its eleven dimensional gravity dual arising from M2 branes. Another class of 3d supersymmetric field theories, which we are interested in here, arises as the low energy limit of  M5 branes wrapping a hyperbolic 3-manifold \cite{Dimofte:2011ju}.  In this manuscript we pursue the counting of microstates for the case of wrapped M5 branes, we will encounter several advantages along the way over M2-brane setups with which the reader might be more familiar. 

The advantages are mainly rooted in the 3d-3d correspondence, which relates the 3d field theory on $N$ wrapped M5's to pure Chern-Simons theory defined on the hyperbolic 3-manifold. This correspondence enables us to calculate the exact $N$ dependence of the partition functions, see \cite{Dimofte:2014ija,Dimofte:2016pua} for reviews on the subject. Most of the results on topologically twisted indices of 3d theories have been obtained at leading order in $N$ \cite{Hosseini:2016tor,Hosseini:2016ume}. One of the key problems plaguing a sub-leading understanding of ${\cal N}=2$ partition functions is that they are given in terms of solutions to certain Bethe-Ansatz equations, and there is no suitable framework to calculate the sub-leading contributions. Indeed,  only a  few partial results for sub-leading structures  have been obtained and they involved substantial numerical efforts   \cite{Liu:2017vll,Liu:2017vbl,Liu:2018bac}. Let us emphasize that going beyond the leading order for the field theory partition function is not merely of academic interest; it  promises to clarify intricate aspects of gravity on the holographic side.

The M5 brane has long been one of the most intriguing and least understood objects in string theory. For M5 branes wrapping a hyperbolic 3-manifold, the holographic description as well as its place  in the context of the 3d-3d correspondence has been elucidated  in a series of works \cite{Gang:2014qla,Gang:2014ema,Gang:2018hjd}. In those works,  the leading large $N$  behavior of supersymmetric quantities, such as the squashed 3-sphere partition function and twisted partition functions, were obtained using 3d-3d relations and matched nicely to the supergravity computations.   In this manuscript we explore subleading corrections to the twisted partition functions, and we present the exact subleading correction terms as the main result. The exact computation  is made possible since 3d-3d relations connect the twisted partition functions to simple topological invariants on 3-manifolds, for which we employ certain mathematical results.  We also provide a gravitational understanding of logarithmic terms in the subleading corrections.

The rest of the manuscript is organized as follows.  We briefly review basic aspects of holography of wrapped M5 branes in section  \ref{Sec:holography}. We cover various entries in the AdS/CFT dictionary and present some of the gravitational backgrounds relevant to our work, we also review the field theory formulation of the relevant partition functions. In section \ref{Sed:PartFuns} we present some of the details of Chern-Simons theory that facilitate the computation of various ingredients in the partition functions  via the 3d-3d correspondence.  Section \ref{Sec:LargeN} develops the large $N$ expansion in some detail.   In section \ref{Sec:Oneloop} we present  the one-loop effective action around the black hole background and show agreement of the $\log N$ terms with the field theory computation.  We conclude in section \ref{Sec:Conclusions} with a summary of our work and by pointing out some interesting open problems. In appendix \ref{app : torsion computation} we present explicit expressions for the analytic torsion used in the main body of the paper.


\section{Holography for wrapped M5-branes}\label{Sec:holography}
In this section we briefly review  the AdS$_4$/CFT$_3$ correspondence associated to wrapped M5-branes on a compact (closed) hyperbolic 3-manifold, $M_3 = \mathbb{H}^3/\Gamma$. To help acquaint the reader with the M5 duality we present a comparison with the more standard form  of AdS$_4$/CFT$_3$  based on  M2-branes probing a cone over a Sasakian-Einstein 7-manifold $Y_7$ in Table \ref{AdS4/CFT3 from M2/M5-branes}. A peculiarity of AdS$_4$/CFT$_3$ from M5-branes is that we can use the  3d-3d correspondence  \cite{Dimofte:2010tz,Terashima:2011qi, Dimofte:2011ju,Terashima:2011xe,Dimofte:2011jd,Cecotti:2011iy,Dimofte:2011py,Dimofte:2013iv,Gang:2013sqa,Yagi2013,Lee2013,Cordova2017,Chung:2014qpa,Dimofte:2014zga,Gukov:2015sna,Pei:2015jsa,Gang:2015wya,Bae:2016jpi,Gukov:2017kmk,Mikhaylov:2017ngi,Gang:2017lsr,Gang:2018wek}
	which  provides an alternative way of computing some supersymmetic quantities, using geometry.   Moreover, this geometrical perspective becomes a conduit to the possibility of exact results in $N$.
\begin{table}[h]
	\begin{center}
		\begin{tabular}{|c|c|c|}
			\hline 
			AdS$_4$/CFT$_3$ & from M2-branes &  from M5-branes
			\\
			\hline \hline
			M-theory set-up& $N$ M2-branes probing Cone($Y_7$)& $N$ M5-branes wrapped on $M_3$ \\ 
			\hline 
			Dual  & Known only for  & Systematic algorithm  \\
			Field theory & special examples of $Y_7$ & applicable to general $M_3$ \\ 
			\hline 
			Gravity dual & $AdS_4 \times Y_7$&  Warped $AdS_4 \times M_3 \times \tilde{S}^4$  \\ 
			\hline 
			Symmetry & Isometry of $Y_7$ ($ \supset  U(1)_R$) & Only  $U(1)_R$ 
			\\
			\hline
			$L^2/G_4$ & $\frac{N^{3/2} \pi^2}{\sqrt{27/8 \textrm{vol}(Y_7)}} $  &   $\frac{2 N^3 \textrm{vol}(M_3)}{3 \pi^2}$
			\\
			\hline
			$L/L_{p}$ & $\propto N^{1/6}$  &   $\propto N^{1/3}$
			\\
			\hline
		\end{tabular} \caption {Comparison between two well-established classes of AdS$_4$/CFT$_3$  using M-theory. $Y_7$ is a Sasakian-Einstein 7-manifold while $M_3$ is a closed hyperbolic 3-manifold. $L_p$ is the Planck length and $L$ is the radius of the $AdS_4$.}
		\label{AdS4/CFT3 from M2/M5-branes}
	\end{center}
\end{table}


On the field theory side, we consider a large class of 3d $\mathcal{N}=2$ superconformal field theories (SCFTs) known as  $\mathcal{T}_{N}[M_3]$ arising from wrapped M5-branes on a compact hyperbolic 3-manifold $M_3$.  
\begin{align}
\begin{split}
&\textrm{$N$ M5-branes : } \;\;\; \mathbb{R}^{1,2} \times M_3\; (\subset T^*M_3)
\\
&\xrightarrow{\textrm{ \;\;\;Low energy worldvolume theory \;\;\; }} \textrm{ 3d $\mathcal{N}=2$ SCFT $\mathcal{T}_N[M_3]$ on $\mathbb{R}^{1,2}$}\;.
\end{split}
\end{align}
The system preserves  4 supercharges and the infra-red (IR) world-volume theory generically has 3d $\mathcal{N}=2$ supersymmetry. 
The field theoretic way to understand this situation is:
\begin{align}
\begin{split}
&\textrm{6d $A_{N-1}$ (2,0) theory on $\mathbb{R}^{1,2}\times M_3$}
\\
&\xrightarrow{\textrm{ \;\;\;Low energy effective theory \;\;\; }} \textrm{ 3d $\mathcal{N}=2$ SCFT $\mathcal{T}_N[M_3]$ on $\mathbb{R}^{1,2}$}\;.
\end{split}
\end{align}
%
To preserve some supersymmetries, we perform a partial topological twisting along $M_3$ using  the $SO(3)$ subgroup of the $SO(5)$ R-symmetry of the 6d theory.  The topolgical twisting preserves 4 supercharges and $SO(2)$ R-symmetry out of the $SO(5)$.
For generic $N$, the  3d theory has only the  $U(1)=SO(2)$ R-symmetry and no other flavor symmetry. Practically, the absence of flavor symmetry implies  that in the partition functions there are no extra fugacities and, therefore, we are limited to the universal sector. This is precisely the situation described in  \cite{Azzurli:2017kxo}, albeit from a different embedding point of view,  as we will discussed below.

\subsection{Holographic dual}

To anticipate details of the holographic description we start in eleven dimensions where the M5-brane naturally resides. Moreover,  to incorporate the 3d hyperbolic manifold $M_3$ we consider its cotangent bundle denoted by $T^*M_3$ which is a local Calabi-Yau. 
\begin{align}
\begin{split}
&\textrm{11d space-time : }  \mathbb{R}^{1,2}\times (T^*M_3)\times \mathbb{R}^2.
\end{split}
\end{align}
Now the holographic background should be the back-reacted AdS solution where M5-branes are partly wrapped on $M_3$.
When $M_3$ is the hyperbolic space, the gravity dual of $\mathcal{T}_{N}[M_3]$ is proposed to be \cite{Gauntlett:2000ng}
\begin{align}
\textrm{AdS$_4$/CFT$_3$ : }  \textrm{$\mathcal{T}_N[M_3]$} \quad  =   \quad  (\textrm{ M-theory on Pernici-Sezgin $AdS_4$ solution }) \;. \label{AdS4/CFT3 for wrapped M5s}
\end{align}

Here Pernici-Sezgin (PS) solutions \cite{Pernici:1984nw} are magneto-vac solutions of 7d $SU(2)$-gauged supergravity, and they include an $AdS_4$ whose 11d uplift is the gravity dual we are looking for.
In 11d, the solution takes the form of a warped product $AdS_4 \times M_3\times \tilde{S}^4$ with 4-form fields turned on along various directions, see {\it e.g.} \cite{Gang:2014ema} for details. The $\tilde{S}^4$ is a squashed 4-sphere with $U(1)$ isometry which corresponds to $U(1)$ R-symmetry in the field theory.

The explicit construction of the gravity solutions exploits the fact that locally the hyperbolic 3-manifold $M_3$ looks like $\mathbb{H}^3$. However, globally we need to consider the quotient $\mathbb{H}^3/\Gamma$. We will explain how the group $\Gamma$ is to be obtained from the construction of the hyperbolic 3-manifold we use in other sections where we consider closed 3-manifold $M_3$ obtained by a Dehn surgery along a knot $K$. For the impatient reader we anticipate that $\Gamma=\pi_ 1(M_3)$.


The maximally supersymmetric 7d $SO(5)$-gauged supergravity is a consistent truncation of 11d supergravity, and in turn it can be again consistently truncated to a 4d $\mathcal{N}=2$ gauged supergravity via a consistent truncation \cite{Donos:2010ax}. The Einstein-graviphoton part of the action is simply
\begin{align}
\label{Eq:4dGrav}
I = \frac{1}{16\pi G_4} \int d^4 x \sqrt{-g} \left( R + \frac{6}{L^2} - \frac{L^2}4 F^2 \right)\;.
\end{align}

This is also the universal sector discussed  recently in  \cite{Azzurli:2017kxo} in the context of microscopic counting of black hole entropy. Here the crucial difference is the embedding into M-theory and, more precisely, the scaling of Newton's constant with the number of branes $N$. 
In the action above $F$ is the field strength for $U(1)$ gauge field in $AdS_4$ which couples to the $U(1)$ R-symmetry in the boundary CFT. The 4d Newton constant $G_4$ after the consistent truncation is related to $N$ in the following way \cite{Gang:2014ema} .
\begin{align}
G_4/L^2 =\frac{3 \pi^2}{2 N^3 \textrm{vol}(M_3)}\;. 
\end{align}

The standard embedding, or the consistent truncation leading to the above pure $\mathcal{N}=2$ supergravity Lagrangian can also be obtained in the context of M2 solutions where $U(1)_R$ is realized geometrically as the Reeb vector of $Y_7$, which is a Sasaki-Einstein space so always written as a $U(1)$ bundle over a K\"ahler-Einstein 6d space \cite{Gauntlett:2007ma}.  In the consistent truncation of \cite{Gang:2014ema}, however,  the $U(1)_R$ direction is identified with an unbroken isometry of the squashed $\tilde{S}^4$.

\subsection{ Twisted partition functions  dual to wrapped M5 branes}

Let us now discuss the field theory dual of some of the entries in the AdS/CFT dictionary pertaining to wrapped M5 branes. We are particularly interested in a certain class of partition functions when the field theory is not placed on the typical $\mathbb{R}^{1,2}$ but on   a more general background. Namely, we are interested in placing the effective 3d field theory on a circle bundle over a genus$-g$ Riemann surface $\Sigma_g$.  The approach to this problem requires that we preserve supersymmetry on curved backgrounds $\mathcal{M}^{\nu_R}_{p,g}$. Later, we will restrict our attention to the case when $p\in 2\mathbb{Z}$ and $\nu_R = \frac{1}2$.

Our goal is to understand the   holographic dictionary using  twisted partition functions $\mathcal{Z}^{\nu_R}_{g,p}$ \cite{Closset:2017zgf,Closset:2018ghr} which are defined on  the curved background $\mathcal{M}_{p,g}$ which denotes a $S^1$-bundle of degree $p$ over a Riemmann surface $\Sigma_g$ of genus $g$, that is, 

\begin{align}
\begin{split}
&  S^1 \xrightarrow{\;  p \;} \mathcal{M}_{g,p} \rightarrow \Sigma_g \;. \label{M-g,p}
\end{split}
\end{align}  
%
The metric is
\begin{align}
ds^2 = \beta^2 \big{(}d \psi - p a(z,\bar{z})\big{)}^2 + 2 g_{z \bar{z}} dz d \bar{z}\;,
\end{align}
where $z,\bar{z}$ are local coordinates on the Riemann surface and $\psi \sim \psi+2\pi$ parameterizes the $S^1$-fiber of length $\beta$. The 1-form  $a$ on $\Sigma_g$  has   curvature $F_a :=da$  normalized as
\begin{align}
\frac{1}{2\pi} \int_{\Sigma_g} d a =1\;.
\end{align} 
To preserve some supersymmetries, we turn on the following background gauge field coupled to $U(1)$ R-symmetry. 
\begin{align}
A^{R}=  \beta \nu_R (d \psi - p a) + n_R (\pi^*  a)\;, \label{background U(1)-R}
\end{align}
with proper quantization conditions for $(\nu_R, n_R)$ \cite{Toldo:2017qsh}. Here $\pi^* a $ is a 1-form on $\mathcal{M}_{g,p}$ given as the pull-back of $a$ using the projection map $\pi : \mathcal{M}_{g,p} \rightarrow \Sigma_g $.  Large gauge transformations relate
\begin{align}
(\nu_R, n_R ) \sim (\nu_R +1 , n_R + p)\;.
\end{align}
%
%
For even $p$, $\mathbb{Z}_2 \subset H_1 (\mathcal{M}_{g,p},\mathbb{Z}_2)$ and we can consider two types of supersymmetric backgrounds with different choice of spin-structures
\begin{align}
p \in 2 \mathbb{Z} \;: \; (\nu_R, n_R) = (0,g-1)  \textrm{ or }   (\frac{1}2,g-1+\frac{p}2)\;.
\end{align} 
The $\nu_R=0$ corresponds to the usual periodic boundary condition while $\nu_R = \frac{1}2$ corresponds to anti-periodic boundary condition under the $\mathbb{Z}_2$. For odd $p$, only the background with $\nu_R =0$ is allowed.  When $p=0$,  the partition functions on the two curved backgrounds have the following interpretation
\begin{align}
\mathcal{Z}^{\nu_R = 0}_{p=0,g} = \textrm{Tr}_{\mathcal{H}(\Sigma_g)} (-1)^{2j_3}\;, \quad \mathcal{Z}^{\nu_R = \frac{1}2}_{p=0,g} = \textrm{Tr}_{\mathcal{H}(\Sigma_g)} (-1)^{R}\;.
\end{align}
Here $j_3$ is the Lorentz spin and $R$ is the charge of $U(1)$ R-symmetry.  We focus on the case $p\in 2\mathbb{Z}_{\geq 0}$ where there are two possible supersymmetric choices of $\nu_R$,  $0$ or $\frac{1}2$, depending on  spin-structure. We further restrict our consideration to the case $\nu_R = \frac{1}2$:
\begin{align}
\nu_R = \frac{1}2\;, \quad n_R = \frac{p}2 + g-1\;, \quad p \in 2 \mathbb{Z}\;. \label{background U(1)-R-2}
\end{align}  
For $p=0$ case, the partition function determined by this background or boundary conditions counts ground states of the 3d theory on a topologically twisted Riemann surface $\Sigma_g$ with signs. 

The twisted  partition functions for general $\mathcal{N}=2$ theory can be written as  \cite{Nekrasov:2014xaa,Closset2017,Closset:2018ghr}
\begin{align}
\mathcal{Z}^{\nu_{R} }_{p,g}  =  \sum_{\alpha}(\mathcal{H}^{\alpha}_{\nu_R})^{g-1} (\mathcal{F}_{\nu_R}^\alpha)^p\;,
\end{align}
where $\alpha$ labels vacua of the 3d $\mathcal{N}=2$ on $\mathbb{R}^2 \times S^1$, called Bethe-vacua, and $\mathcal{H}$ and $\mathcal{F}$ are called `handle-gluing' and `fibering' operators, respectively.

\subsection{Taub-Bolt solutions in $AdS_4$}
According to the standard dictionary of AdS/CFT, the twisted partition functions at leading order  in the $1/N$ expansion can be  holographically computed from the on-shell gravitational action  
\begin{align}
\mathcal{Z}^{\nu_{R} = \frac{1}2 }_{p,g}  (\mathcal{T}_N[M_3]) = \sum_{\hat{\alpha}} e^{-I^{\rm gravity}_{p,g}(\hat{\alpha})}\;.
\end{align}
%
Here $\hat{\alpha}$ runs over all the large $N$ saddle points of the M-theory 
which asymptotically approach the $\mathcal{M}^{\nu_R = \frac{1}2}_{p,g}$ geometry in the  AdS$_4$ boundary.
In recent work \cite{Toldo:2017qsh}, two BPS supergravity solutions called Taub-Bolt solutions (Bolt$_{\pm}$) with asymptotic boundary $\mathcal{M}^{\nu_R = \frac{1}2}_{p \in 2\mathbb{Z},g}$  were constructed. From the computation of the holographically renormalized on-shell supergravity actions for the two solutions, we have
\begin{align}
\begin{split}
I^{\rm gravity}_{p,g} (\textrm{Bolt}_\pm) &= \frac{\pi (4(1-g)\mp p) L^2}{8G_4} +(\textrm{subleading corrections in $G_4$}) \;,
\\
& = \frac{ (4(1-g)\mp p)N^3}{12 \pi} \textrm{vol}(M_3) +(\textrm{subleading corrections in $1/N$}) \;. \label{leading-I-bolt}
\end{split}
\end{align}
For the sub-leading  corrections, we need to consider  M-theory on the 11d uplifted supergravity background.  Assuming that the $\textrm{Bolt}_+$ (and $\textrm{Bolt}_-$) solution gives a dominant contribution in the large $N$ limit for $p>0$ ($p=0$),  holography predicts
\begin{align}
\begin{split}
&\mathcal{Z}^{\nu_{R} = \frac{1}2 }_{p \geq  0,g}  (\mathcal{T}_N[M_3]) 
\\
&\xrightarrow{ \quad N \rightarrow \infty \quad }
\exp \left( \frac{(4(g-1)+p)N^3}{12\pi} \textrm{vol}(M_3) + \textrm{subleading}\right) \left(1+e^{-(\ldots) }\right)\;.
\end{split}
\end{align}
Here $e^{- (\ldots)}$  stands for  exponentially suppressed terms at large $N$.

\subsection{Magnetically charged  black hole in $AdS_4$ }
For $p=0$, the twisted  partition functions have an alternative interpretation on the  holographic dual side. It counts the microstates of a magnetically charged black hole  in $AdS_4$  with signs.   

The gauged supergravity admits the  following $1/2$ BPS magnetically charged asymptotically $AdS_4$ black hole solution\cite{Romans:1991nq,Caldarelli:1998hg,Chamseddine:2000bk,Cacciatori:2009iz}
\begin{align}
\begin{split}
\label{Eq:BlackHole}
&\frac{ds^2}{L^2}= -(\rho  - \frac{1}{2\rho})^2 dt^2 + \frac{1}{(\rho - \frac{1}{2\rho})^2} d \rho^2 + \rho^2 ds^2 (\Sigma_g)\;,
\\
&F =\frac{1}{L^2} (\textrm{volume form on $\Sigma_g$})\;.
\end{split}
\end{align}
Thanks to the consistent truncation discussed previously, any solution of the 4d action above can be embedded in 11d in a way that admits an M5 brane  interpretation.  Note that this is the starting point of \cite{Azzurli:2017kxo}, which considered what we call the M2 embedding into 11d supergravity. 

We assume $g>1$ and $ds^2 (\Sigma_g)$ is a uniform hyperbolic metric on a Riemann surface $\Sigma_g$ of genus $g$ normalized as $\textrm{vol}(\Sigma_g) = 4\pi (g-1)$. The black hole solution  interpolates asymptotically $AdS_4$ with    conformal boundary $\mathbb{R}_t\times \Sigma_g$ and  $AdS_2 \times \Sigma_g$ near-horizon geometry:
\begin{align}
\begin{split}
AdS_4\; (\rho = \infty) \rightarrow AdS_2 \times \Sigma_g \;(\rho = \frac{1}{\sqrt{2}})
\end{split}
\end{align}
The Bekenstein-Hawking entropy of the black hole  is 
\begin{align}
\begin{split}
S_{\rm BH} &= \frac{A_{\rm horizon}}{4 G_4} + (\textrm{subleadings})=  \frac{2\pi (g-1) L^2}{4G_4} + (\textrm{subleadings}) \;,
\\
&= \frac{(g-1)\textrm{vol}(M_3)N^3}{3\pi} + (\textrm{subleadings})\;. \label{leading SBH}
\end{split}
\end{align}
Ultimately, the entropy should be understood from  `microstates counting' of the asymptotically $AdS_4$ black hole:
\begin{align}
S_{\rm BH} (g,N,M_3)=\log  d(g,N,M_3).
\end{align}
Via the AdS$_4$/CFT$_3$ correspondence, the number of  black hole microstates $d(g,N,M_3)$ is mapped to the number of ground states on $\Sigma_g $ in the dual $\mathcal{T}_N[M_3]$.
\begin{align}
\begin{split}
\textrm{AdS$_4$/CFT$_3$ : } d(g,N,M_3) & :=d^{R \in 2\mathbb{Z}}(g,N,M_3) +d^{R \in 2\mathbb{Z}+1}(g,N,M_3) \;.
\\
&= (\textrm{the number of ground states of $\mathcal{T}_N [M_2]$ on $\Sigma_g$})\;.
\end{split}
\end{align}
On the other hand, the twisted partition function  for $p=0$ computes 
\begin{align}
\mathcal{Z}^{\nu_R = \frac{1}2}_{p=0,g} = d^{\rm SUSY} (g,N, M_3):= d^{R \in 2\mathbb{Z}}(g,N,M_3) -d^{R \in 2\mathbb{Z}+1}(g,N,M_3)\;.
\end{align}
Although there could be huge cancellations between states with $R \in 2\mathbb{Z}$ and $R\in 2\mathbb{Z}+1$, the twisted index  turns out to reproduce the Bekenstein-Hawking entropy of AdS${}_4$ black hole at large $N$ for various models of AdS$_4$/CFT$_3$ \cite{Benini:2015eyy,Benini:2016rke,Cabo-Bizet:2017jsl,Benini:2017oxt,Hosseini:2017fjo,Azzurli:2017kxo}.  It would be interesting to see if they also match even at finite $N$.  One of the goals of this paper is to explore the microstate counting beyond the leading order on the field theory side and subsequently perform a one-loop effective action calculation on the gravity side that leads to the logarithmic in $N$ term.


\section{3d-3d relation for  twisted partition functions}\label{Sed:PartFuns}
In this section, we relate the twisted partition functions  $\mathcal{Z}^{\nu_R = \frac{1}2}_{g,p\in 2 \mathbb{Z}}(\mathcal{T}_N[M_3])$ to simple topological quantities on the 3-manifold $M_3$ via a 3d-3d relation. The final expression is given in \eqref{dmicro-sum-over-flat-connetions}.   We derive the 3d-3d dictionary using the explicit field theoretic construction of $\mathcal{T}_{N}[M_3]$ proposed in \cite{Dimofte:2011ju,Dimofte:2013iv,Gang:2018wek}. As an non-trivial consistency check, we confirm the integrality of the topological quantities for the $p=0$ case with explicit examples.

The relevant 3d-3d relations are summarized in Table~\ref{3d/3d correspondence for twisted index}.
\begin{table}[h]
	\begin{center}
		\begin{tabular}{|c|c|}
			\hline 
			3D $\mathcal{T}_{N}[M_3]$ theory on $\mathbb{R}^2 \times S^1$ & $PSL(N, \mathbb{C})$ CS theory on $M_3$ \\
			\hline \hline
			Bethe vacuum  $\alpha$ & \;\; Irreducible  flat connection $\mathcal{A}^{\alpha} $\;\; \\ 
			\hline 
			Fibering  operator $\mathcal{F}^{\alpha}_{\nu_R = \frac{1}2}$ & $\exp (-\frac{1}{2\pi i }S^\alpha_0)=\exp (\frac{1}{4\pi i }CS[\mathcal{A}^\alpha;M_3]) $  \\ 
			\hline 
			Handle gluing operator $\mathcal{H}^{\alpha}_{\nu_R = \frac{1}2}$ & $ N \exp(-2S_1^\alpha) =N \times {\bf Tor}_{M_3}^{(\alpha)} (\tau_{\rm adj},N) $  \\ 
			\hline 
		\end{tabular} \caption {A 3d-3d dictionary for basic ingredients in twisted ptns  computation.
			$\mathbf{Tor}^{(\alpha)}_{M_3} (\tau,N)$ is analytic torsion (Ray-Singer torsion) for an associated vector bundle in a representation $\tau \in \textrm{Hom}[PSL(N,\mathbb{C}) \rightarrow {GL}(V_\tau)]$  twisted by a flat connection $\mathcal{A}^{\alpha}$. The dictionary for the handle gluing operator works only for $M_3$ with vanishing $H_1 (M_3, \mathbb{Z}_N)$.  }
		\label{3d/3d correspondence for twisted index}
	\end{center}
\end{table}
The Chern-Simons functional is
\begin{align}
CS[\mathcal{A};M_3] := \int_{M_3} \textrm{Tr}(\mathcal{A}d \mathcal{A} + \frac{2}3 \mathcal{A}^3)
\end{align}
Extremizing the functional, we have flat-connection  equation
\begin{align}
d \mathcal{A} + \mathcal{A} \wedge \mathcal{A} = 0\;.
\end{align}
There is a one-to-one correspondence between
\begin{align}
\begin{split}
&\{ \textrm{$PSL(N,\mathbb{C})$ flat-connections $\mathcal{A}^{\alpha}$ on $M_3$} \}/(\textrm{gauge equivalence}) \; 
\\
& \xleftrightarrow{\rm  \;\; 1-1 \;\; } \{ \rho_{\alpha} \;:\; \rho_\alpha \in \textrm{Hom}[\pi_1 (M_3)\rightarrow PSL(N,\mathbb{C})]/(\rm \textrm{conjugation}) \}.
\end{split}
\end{align}
For $N=2$, the gauge $PSL(N,\mathbb{C})$ is identical  to the orientation-preserving isometry group of hyperbolic upper half-plane $\mathbb{H}^3$. 
\begin{align}
\textrm{Isom}^{+}(\mathbb{H}^3) = PSL(2,\mathbb{C})\;.
\end{align}
For hyperbolic $M_3$, it is known that there is a discrete and faithful $PSL(2,\mathbb{C})$ representation $\rho_{\rm geom}$
\begin{align}
\rho_{\rm geom} :  \textrm{  discrete and faithful $PSL(2,\mathbb{C})$ representation of $\pi_1 (M_3)$}\;. \label{discrete-faithful rep}
\end{align}
The representation $\rho_{\rm geom}$ above precisely furnishes a geometric construction of the 3-manifold as $M_3=\mathbb{H}^3/\Gamma$ where the group action $\Gamma$ is identified with
\begin{align}
\Gamma = \rho_{\rm geom}\left(\pi_1 (M_3)\right) \subset \textrm{Isom}^+ (\mathbb{H}^3)\;.
\end{align}
$\textrm{\bf Tor}^{\alpha}_{M_3} (\tau,N)$ denotes the analytic torsion \cite{ray1971r} of an associated vector bundle for $PSL(N,\mathbb{C})$ principal bundle over $M_3$ in a representation $\tau \in \textrm{Hom}\left(PSL(N,\mathbb{C}) \rightarrow GL(V_\tau)\right)$\footnote{The torsion here is  the inverse of torsion in some mathematical literatures. For example,  $\textrm{\bf Tor}_{M_3}(\tau)$  is $1/T_{M_3}(\tau)$ in \cite{muller2012asymptotics}.}
\begin{align}
{\bf Tor}_{M_3}^{(\alpha)}  (\tau,N) := \frac{[\det' \Delta_1 (\tau, \mathcal{A}^\alpha)]^{1/2}}{[\det' \Delta_0 (\tau, \mathcal{A}^\alpha)]^{3/2}}\;.
\end{align}
Here $\Delta_n (\tau, \mathcal{A}^\alpha)$ is a Laplacian action on $V_\tau$-valued $n$-form twisted by a $PSL(N,\mathbb{C})$ flat connection $\mathcal{A}^\alpha$:
\\
\begin{align}
\Delta_n (\tau, \mathcal{A})  = d_A * d_A * + *d_A *d_A \;, \quad d_A  = d+ \mathcal{A} \wedge_\tau\;.
\end{align}
Note that $d_A^2 =0$ for  flat connections $\mathcal{A}$.  In Table \ref{3d/3d correspondence for twisted index}  $\tau_{\rm adj}$  denotes the adjoint representation of $PSL(N,\mathbb{C})$. In the above, $\det' \Delta_n$ is the zeta  regularized determinant of the Laplacian $\Delta_n$.
The torsion for  adjoint representation, $\tau = \tau_{\rm adj}$, is related to the one-loop perturbative correction of  $PSL(N,\mathbb{C})$ Chern-Simons theory \cite{witten1989quantum,Gukov:2006ze}
\begin{align}
\begin{split}
&\int \frac{D(\delta \CA)}{(\rm gauge)} e^{-\frac{1}{2\hbar} CS[\mathcal{A}^\alpha +\delta \CA ;M_3]} 
\\
&\xrightarrow{\quad \hbar \rightarrow 0 \quad }  \exp \left(\frac{1}{\hbar}S_0^\alpha + S_1^\alpha + o(\hbar)\right) 
\\
&\propto \exp \left({-\frac{1}{2\hbar} CS[\mathcal{A}^\alpha  ;M_3]}\right) \frac{1}{\sqrt{{\bf Tor}^{\alpha}_{M_3} (\tau_{\rm adj}, N)}} \big{(}1+ o (\hbar) \big{)}\;. \label{perturbative expansion of CS ptn}
\end{split}
\end{align}
In the expansion, we are sloppy in the subtle overall factor independent on $\hbar$ and use the symbol `$\propto$' instead of `$=$'. Using the relations, we finally have
\begin{align}
\begin{split}
&\mathcal{Z}^{\nu_R = \frac{1}2}_{p \in 2\mathbb{Z},g} (\mathcal{T}_N[M_3]) =  \sum_{ \alpha \in \chi^{\rm irred}(N,M_3) } \exp \left(p \frac{CS[\mathcal{A}^\alpha]}{4\pi i } \right) N^{g-1}\left( {\bf Tor}_{M_3}^{(\alpha)}  (\tau_{\rm adj},N)\right)^{g-1}\;,
\\
& \chi^{\rm irred}(N,M_3)  = \{\textrm{set of irreducible $PSL(N,\mathbb{C})$ flat-connections on $M_3$}\} \label{dmicro-sum-over-flat-connetions}\;,
\end{split}
\end{align}
for arbitrary closed hyperbolic 3-manifold $M_3$ with vanishing $H_1 (M_3,\mathbb{Z}_N)$.

\subsection{Derivation}
Some ingredients of the 3d-3d dictionary were originally studied in \cite{Dimofte:2010tz}. In that work, a one-to-one correspondence between $PSL(N,\mathbb{C})$ flat connections on $M_3$ and Bethe vacua on $\mathbb{R}^2\times S^1$ was found. In the correspondence, the on-shell  twisted superpotential of a Bethe vacuum is identified with the classical $PSL(N,\mathbb{C})$ Chern-Simons action of the corresponding flat connection. Combined with the fact that $\mathcal{T}_N [M_3]$ does not have any flavor symmetry, it explains the entry in the  dictionary in  Table \ref{3d/3d correspondence for twisted index} for fibering operators modulo a subtle issue which we now discuss. The issue is  whether all flat connections are relevant for the 3d theory $\mathcal{T}_N[M_3]$, or only a subset is enough. 

\paragraph{$\mathcal{T}^{\rm full}_N[M_3]$ versus $\mathcal{T}^{DGG}_N[M_3]$:} The subtle issue becomes more relevant in 3d-3d correspondence after a concrete and beautiful field theoretic construction, say $\mathcal{T}^{\rm DGG}_N[\mathbb{N}]$,  for 3-manifolds $\mathbb{N}$ with torus boundaries was proposed in \cite{Dimofte:2011ju}.  The construction is based on an ideal triangulation of $\mathbb{N}$ and thus can not see all the flat connections on $\mathbb{N}$ but only sees irreducible flat connections \cite{Chung:2014qpa}.  The construction has been generalized to the case of closed 3-manifold $M_3$ (without any boundary),  say $\mathcal{T}^{\rm DGG}_N[M_3]$, in \cite{Gang:2018wek} by incorporating Dehn filling operation to  the Dimofte-Gaiotto-Gukov's construction. The construction for closed 3-manifold also can not see reducible flat connections. Taking the absence of reducible flat connections as a serious problem, it is argued that there should be a better, alternative field theoretic construction, say $\mathcal{T}^{\rm full}_N[M_3]$, which contains all the flat connections on $M_3$ as Bethe vacua on $\mathbb{R}^2\times S^1$  \cite{Chung:2014qpa}.  Later, concrete field theoretic descriptions of $\mathcal{T}^{\rm full}_{N}[M_3]$ for certain classes of  non-hyperbolic 3-manifolds are proposed in \cite{Pei:2015jsa,Gukov:2017kmk}.\footnote{For non-hyperbolic 3-manifolds $M_3$, on the other hand, the corresponding $\mathcal{T}^{DGG}_{N}[M_3]$ theories are rather trivial, either  mass gapped topological theory (possible with decoupled free chirals)  or a theory with spontaneously broken supersymmetry.} But, as far as we are aware of, there is no known concrete example of $\mathcal{T}_{N}^{\rm full}[M_3]$ for hyperbolic $M_3$.\footnote{One tricky example is the case when  3-manifolds are  mapping tori over one-punctured torus with $N=2$ as studied in \cite{Terashima2011,Gang:2013sqa}. On the 3-manifolds, there exist reducible $PSL(2,\mathbb{C})$ flat-connections only when the $PSL(2,\mathbb{C})$ holonomy around the puncture is trivial. The eigenvelues of the puncture holonomy is related to the real mass parameter coupled to a $U(1)_{\rm punt}$ flavor symmetry in the corresponding  $\mathcal{N}=2$ field theory. In general, only the Bethe-vacua at generic values of real mass parameters of a 3d gauge theory have a definite physical meaning,  i.e.  invariance under IR dualities. Ignoring the unphysical reducible flat connections, there are only irreducible $PSL(2,\mathbb{C})$ flat connections on the 3-manifold and thus $\mathcal{T}^{\rm full} = \mathcal{T}^{DGG}$. See the section 4.1.1 of \cite{Dimofte:2014ija} for the similar story for mapping cylinder case.  }  This is rather surprising and disappointing  since most 3-manifolds are  hyperbolic \cite{thurston1979geometry}. If one only wishes to see irreducible flat connections, then the 3d theories corresponding to  small hyperbolic 3-manifolds can be easily identified \cite{Gang:2017lsr}.   
More recently, the subtle issue was revisited in \cite{Gang:2018wek} where it was argued that for hyperbolic 3-manifolds $M_3$ we do not expect to see all flat connections from a single 3d effective theory $\mathcal{T}_N[M_3]$.  This is because for general hyperbolic 3-manifold, there can be several disconnected components in the vacuum moduli space on $\mathbb{R}^3$ of the 6d twisted compactification along $M_3$.  Thus, we need to choose a single  branch in taking the low energy limit and we  only see the single branch in the effective low-dimensional theory. If this argument is correct, the existence of non-trivial superconformal field theory $\mathcal{T}^{DGG}_N[M_3]$ (which only sees irreducible flat connections) implies the  {\it non-existence of $\mathcal{T}^{\rm full}_N[M_3]$ for hyperbolic $M_3$}.\footnote{Here, let us speculate on why we can construct  $\mathcal{T}^{\rm full}_N[M_3]$  for non-hyperbolic manifolds studied in \cite{Pei:2015jsa,Gukov:2017kmk}. In that cases, the twisted compactification of 6d $(2,0)$ theory enjoys an additional flavor symmetry, say $U(1)_\beta$, due to a Seifert-fibered structure on the 3-manifolds. Thanks to the additional symmetry, we can introduce a suspersymmetry preserving real mass deformation and the continuous deformation may connect all the Bethe-vacua on $\mathbb{R}^2\times S^1$ of the system in a way that the vacua on $\mathbb{R}^3$ after the decompactification    has a single component. This  might be the reason why we can see all flat-connections from a single effective 3d gauge theory. In 3d-3d correspondence, partition functions of $\mathcal{T}_{N}[M_3]$ on squashed Lens spaces are identified with  partition functions of $PSL(N,\mathbb{C})$ Chern-Simons theories on $M_3$ \cite{Dimofte:2014zga}. Reducible flat connections can not contribute  to the complex CS partition functions, since its stabilizer group is non-compact with infinite volume. This is a crucial difference between Chern-Simons theory with compact and non-compact gauge group.  For non-hyperbolic  3-manifolds considered in \cite{Pei:2015jsa,Gukov:2017kmk}, one can regularize the infinite volume by turning on real mass, or fugacity, coupled to $U(1)_\beta$ and can see the contributions from reducible flat connections after the regularization. See the section 3 of \cite{Dimofte:2016pua} for more explanations on the point. This is  compatible with our speculation that  the contributions from {\it reducible flat connections} can be seen in a single effective 3d theory  {\it only for 3-manifolds with extra structure} which gives an additional  flavor symmetry $U(1)_\beta$. } 
According to  \cite{Gang:2018wek}, the $\mathcal{T}^{DGG}_N[M_3]$ for hyperbolic $M_3$ is proposed to be the 3d  effective theory sitting on a vacuum on $\mathbb{R}^3$ which becomes the `irreducible' Bethe-vacua on $\mathbb{R}^2\times S^1$ in the compactification  $\mathbb{R}\rightarrow S^1$. The proposal has been supported by various independent reasonings, such as the resurgence analysis \cite{Gang:2017hbs}
and explicit field theoretic checks \cite{Gang:2018wek,Gaiotto:2018yjh,Benini:2018bhk,Gang:2018huc} of the symmetry enhancements of $\mathcal{T}^{DGG}_{N}[M_3]$ theories, which are geometrically  predicted from the proposal.  We will assume that the $\mathcal{T}^{DGG}_N[M_3]$ is actually the 3d  theory $\mathcal{T}_{N}[M_3]$ appearing the AdS$_4$/CFT$_3$ correspondence in \eqref{AdS4/CFT3 for wrapped M5s} for hyperbolic $M_3$. 
\begin{align}
\textrm{Basic assumption : } \mathcal{T}^{DGG}_N[M_3] =\left(\mathcal{T}_{N}[M_3] \textrm{ in eq.~\eqref{AdS4/CFT3 for wrapped M5s}}\right) \;.
\end{align}
This assumption  has passed  large $N$ consistency checks  using a squashed 3-sphere partition function \cite{Gang:2014ema}. From now on, we will erase the superscript `DGG' and  derive the dictionary for the handle-gluing operator in Table~\ref{3d/3d correspondence for twisted index} using the explicit field theoretic description.

\paragraph{Sketch of the derivation:} The 3d-3d dictionary for handle gluing operators in Table \ref{3d/3d correspondence for twisted index} follows from direct comparison between localization computation using the explicit field theoretic construction \cite{Dimofte:2011ju,Dimofte:2013iv,Gang:2018wek} of $\mathcal{T}_{N}[M_3]$ and the computation of  $\mathbf{Tor}^{(\alpha)}_{M_3} (\tau_{\rm adj},N)$ using a state-integral model.    
The comparison can be summarized by following diagram:
\begin{displaymath}
\xymatrix{
	M_3 = \left( \bigcup_{i=1}^k \Delta_i \bigcup_{a=1}^s \mathbf{S}_a \right)/\sim \ar[d]^{DGG} \ar[dr]^{\qquad \textrm{state-integral model}}& \\ 
	\textrm{\qquad \qquad\qquad \qquad \; $\mathcal{T}_N[M_3] \xrightarrow[]{\text{  localization  }}  \mathcal{H}^\alpha$ \quad \quad} & \textrm{${\bf Tor}^{(\alpha)}_{M_3}$}}
\end{displaymath}
For a given  closed hyperbolic 3-manifold, we can decompose it into to basic building blocks, {\it i.e.} ideal tetrahedron $\Delta$ and solid-torus $\mathbf{S}$. The topological gluing datum $\sim$ encodes the field theoretic description of $\mathcal{T}_N[M_3]$, and $\mathcal{F}$ and  $\mathcal{H}$ can be computed using the general localization results. From the gluing datum, on the other hand, state-integral models for $PSL(N,\mathbb{C})$ Chern-Simons partition function are developed  and the perturbative invariants, $S^\alpha_0$ and $S^\alpha_1$ in \eqref{perturbative expansion of CS ptn}, can be computed from the state-integral models \cite{2007JGP,Dimofte:2009yn,Dimofte:2012qj,Garoufalidis:2013upa,Dimofte:2013iv,Bae:2016jpi,Gang:2017cwq}. The 1-loop part $S_1^\alpha$ is simply related to the torsion ${\bf Tor}^{(\alpha)}_{M_3}$ as in \eqref{perturbative expansion of CS ptn}.
The comparison is almost straightforward. Our explanation will focus on two subtle points in the 3d-3d dictionary: i) the {\it factor $N$} in the hand-gluing operator $\mathcal{H}$ and ii) the reason why we assume the topological condition $H_1 (M_3, \mathbb{Z}_N) =0$. These subtle issues play important roles  in a) checking  {\it integrality} of the twisted indices at finite $N$ and b) reproducing correct {\it subleading $\log N$ corrections} of twisted indices compatible with supergravity analysis at large $N$. For general hyperboilc 3-manifold $M_3$ with non-vanishing $H_1 (M_3, \mathbb{Z}_N)$, we know the 3d-3d dictionary should be modified slightly but it is not clearly exactly how. 

Some of previous studies on twisted indices in 3d-3d correspondence can be found in \cite{Gukov:2016gkn,Gukov:2017kmk,Gang:2018hjd}. In particular, in \cite{Gang:2018hjd}, two of the current authors proposed an analogous dictionary for handle-gluing operator. The derivation there simply follows from the combination of two known facts, a) a 3d-3d dictionary for perturbative expansions of holomorphic blocks \cite{Dimofte:2010tz,Beem:2012mb} and b) the general relation \cite{Closset:2018ghr,to-appear} between the first two terms in the perturbative expansion and the two operators, $\mathcal{F}_{\nu_R = 1/2}$ and $\mathcal{H}_{\nu_R= 1/2}$. In the derivation, however, there are several subtle issues such as  i) whether the gauge group of complex Chern-Simons theory is $SL(N,\mathbb{C})$ or $PSL(N,\mathbb{C})$ and ii) what is the correct $N$-dependent overall factor in the perturbative expansion \eqref{perturbative expansion of CS ptn}. These subtle issues are irrelevant in computing large $N$ leading behavior of the twisted partition functions.  Since we are now more interested in subleading $1/N$ corrections, we need to be extremely careful in the derivation. So, we will derive the 3d-3d relation directly from the field theoretic construction of $\mathcal{T}_N[M_3]$ without relying on indirect relations. From an honest derivation, we clarify two subtle points mentioned above which were not addressed in \cite{Gang:2018hjd}.

\paragraph{Brief review of the construction of $\mathcal{T}_N [M_3]$:} For simplicity,  consider hyperbolic 3-manifolds  obtained by an integral Dehn surgery along a hyperbolic knot $K$ with a slope $P$.
\begin{align}
M_3 = (S^3\backslash K)_{P\mu + \lambda}\;. 
\end{align}
Refer to appendix \ref{app : torsion computation} for the notation of surgery representation of 3-manifold.  For the field theoretic description for the $\mathcal{T}_N [M_3]$, we first need to consider the 3d theory $\mathcal{T}_N[S^3\backslash K;\mu ]$ associated with the knot complement constructed in \cite{Dimofte:2011ju,Dimofte:2013iv}. The case when $N=2$ was first studied in \cite{Dimofte:2011ju} based on an ideal triangulation of the knot complement
\begin{align}
S^3\backslash K = \left( \bigcup_{i=1}^k \Delta_i\right)/\sim\;.
\end{align}
For an arbitrary hyperbolic knot $K$, an ideal triangulation of the knot complement $S^3\backslash K$ is available in a computer program Snap{P}y \cite{SnapPy}. In \cite{Garoufalidis:2013upa,Dimofte:2013iv}, the construction is generalized to higher $N$  by introducing $N$-decomposition which replace each ideal tetrahedron $\Delta$ in the triangulation by $\frac{1}6 N (N^2-1)$ copies of octahedra $\Diamond$:
\begin{align}
\textrm{$N$-decomposition : $S^3\backslash K = \left( \bigcup_{i=1}^k \Delta_i\right)/\sim \; \longrightarrow\;  \left( \bigcup_{i=1}^k \bigcup_{\alpha=1}^{\frac{1}6 N(N^2-1)} \Diamond^{(\alpha)}_i\right)/\sim$ }
\end{align}
The 3d theory constructed from the $N$-decomposition has an explicit UV field theoretic description whose gauge group is
\begin{align}
U(1)^{\frac{k}6 N (N^2-1)}\;.
\end{align}
There are also as many as $\frac{k}6 N (N^2-1)$ chiral fields in the theory $\mathcal{T}_N [S^3\backslash K, \mu]$ and mixed Chern-Simons levels of the gauge group and superpotential interactions are determined by the $N$-decomposition.  The theory has manifest $U(1)^{N-1}$ flavor symmetry associated to the torus boundary of the knot complement.  In \cite{Gang:2018wek}, it was argued that the UV symmetry is enhanced to $PSU(N)= SU(N)/\mathbb{Z}_N$ in IR.\footnote{The theory $\mathcal{T}_N^{DGG}[S^3\backslash K;\gamma ]$ depends on the choice of primitive  boundary  1-cycle $\gamma \in H_1\left(\partial (S^3\backslash K), \mathbb{Z}\right)$. For generic choice of $\gamma$, there is no symmetry enhancement. When $\gamma$ is chosen such that $\chi^{\rm irred}\left((S^3\backslash K)_\gamma , N \right)$ is empty, the $U(1)^{N-1}$ is  enhanced to $SU(N)$ (or $PSU(N)$) if $\gamma$ is a trivial (or a non-trivial) element in  $H_1 \left( S^3\backslash K, \mathbb{Z}_2 \right) = \mathbb{Z}_2 $. For $\gamma = \mu$ (meridian) case, $\chi^{\rm irred} \left((S^3\backslash K)_\gamma = S^3 , N\right) $ is obviously empty since there is no irreducible flat connections on $S^3$. Note also that  $\mu$ is a generator of $H_1 \left( S^3\backslash K, \mathbb{Z}_2 \right)$.} The 3d gauge theory associated to the Dehn filled closed manifold can be simply obtained by gauging the IR $PSU(N)$ flavor symmetry\footnote{The Chern-Simons level for background gauge field coupled to the $PSU(N)$ symmetry of  $\mathcal{T}_N [S^3\backslash K, \gamma=\mu]$ depends on the choice of dual bounday 1-cycle $\gamma_{\rm dual} \in H_1 \left(\partial (S^3\backslash K) ,\mathbb{Z}\right)$ which intersects  $\gamma$ once. We choose the $\gamma_{\rm dual}$ as longitude $\lambda$. In the gauging, we introduce additional CS interaction of level $P$ in addition to the CS level determined by the choice of $\gamma_{\rm dual}$.}
\begin{align}
\begin{split}
&\mathcal{T}_{N}[M_3=(S^3\backslash K)_{P\mu + \lambda}] 
\\
&= (\textrm{Gauging $PSU(N)$ flavor symmetry of $\mathcal{T}_N[S^3\backslash K;\mu]$ with additional CS level $P$})\;. \label{T[M] under Dehn filling}
\end{split}
\end{align}

\paragraph{Handle-gluing operator for knot complement theory} Applying the  general localization formula to the explicit field theoretic description of $\mathcal{T}_N[S^3\backslash K;\mu]$, it is  straightforward to check that
\begin{align}
\mathcal{H}_{\nu_R = \frac{1}2} (\mathcal{T}_N[S^3\backslash K;\mu])  =\det \left(A(\vec{\mathbb{E}}_{\bf m} )\cdot \Delta_{z''} + B(\vec{\mathbb{E}}_{\bf m}) \cdot \Delta_{z^{-1}} \right) \prod_{}(z_i)^{f''_i} (z''_{i})^{f_i} \label{factor N-1}
\end{align}
%
%
%
Here $A,B$ are square matrices of size $\frac{k}{6} N(N^2-1)$ and $\vec{f}, \vec{f}''$ are vectors of size $\frac{k}{6} N(N^2-1)$. They are determined by the gluing rule of octahedra  in the $N$-decomposition. In the field theory side, the matrices determine the mixed Chern-Simons levels among $U(1)^{k \frac{N(N^2-1)}6 }$  gauge group and the vectors determine the mixed  CS levels between the gauge group and the background $u(1)_R$ gauge field. In the above, we define 
\begin{align}
\begin{split}
&\Delta_{z''} := \textrm{diag} \{z_1'',\ldots, z''_{\frac{k N(N^2-1)}6} \}\;, \quad z''_i := 1- z^{-1}_{i}\;,
\\
& \Delta_{z^{-1}} := \textrm{diag} \{z_1^{-1},\ldots, z_{\frac{k N(N^2-1)}6}^{-1} \}\;.
\end{split}
\end{align}
Here $\{z_i\}$ are exponentiated complexified  holonomy variables along $S^1 \subset \mathbb{R}^2 \times S^1$ for $u(1)^{k \frac{N(N^2-1)}6 }$ gauge group.  The determinant factor comes from the Hessian of the twisted superpotential and remaining products come from the so-called effective dilaton. Bethe vacua are given as solutions of the following algebraic equations
\begin{align}
&\textrm{Bethe equations : }\prod \big{(} (-1)^{f_j}z_j \big{)}^{A_{ij}} \big{(}(-1)^{f''_j}z''_j\big{)}^{B_{ij}}  = \begin{cases}
m_i\;, \quad \textrm{for $1 \leq  i \leq N-1$}\\
1\;, \quad \textrm{for $  i \geq N $} \label{Bethe-equations}
\end{cases} 
\end{align}
Here $\{m_a\}_{a=1}^{N-1}$ are  the  background $S^1$-holonomies copuled to $U(1)^{N-1}$ flavor symmetry of the  $\mathcal{T}_N[S^3\backslash K;\mu]$ theory.  On the 3-manifold side, on the other hand, the $\{z_i\}$ parametrize the shape of octahedra in the $N$-decomposition. The above Bethe equations are actually  identical  to the gluing equations for the  octahedra. For each solution to the gluing equations, there is a corresponding $PSL(N,\mathbb{C})$ flat connection on $S^3\backslash K$ whose holonomy along the boundary meridian cycle $\mu$ is given by
\begin{align}
\begin{split}
&P \exp \left(-\oint_{\mu \in H_1 \left(\partial (S^3\backslash K ),\mathbb{Z}\right)} \mathcal{A} \right) = \exp \left(\sum_{a=1}^{N-1}\mathbb{E}^a \log m_a(\vec{z},\vec{z}'')\right)\;. \label{boundary holonomy along meridian}
\end{split}
\end{align}
The $m_a$  depends on the  choice $\{\mathbb{E}^a\}$ of $psu(N)$ Lie-algebra basis,  such that $\sum \mathbb{E}^a \log m_a$ is kept invariant.  Through the Bethe equations in \eqref{Bethe-equations}, the first $(N-1)$ rows, say $(A_{N-1},B_{N-1})$, of the  matrices $(A,B)$  also depend on the choice.  Under the basis change $\vec{\mathbb{E}}_1 \rightarrow \vec{\mathbb{E}}_2 = g\cdot \vec{\mathbb{E}}_1$ with $g\in GL(N-1, \mathbb{R})$, the matrices $A_{N-1}$ and $B_{N-1}$ transform as
\begin{align}
\begin{split}
&A_{N-1} (\vec{\mathbb{E}}_2) = (g^{-1})^T \cdot A_{N-1}(\vec{\mathbb{E}}_1)\;,
\\
&B_{N-1} (\vec{\mathbb{E}}_2) = (g^{-1})^T \cdot B_{N-1}(\vec{\mathbb{E}}_1)\;. 
\label{factor N-3}
\end{split}
\end{align}
For the resulting 3d theory to have properly quantized mixed CS levels, the matrices should be integer valued.  Upon  the following choice of basis  $\{\vec{\mathbb{E}}_{\bf m}\}$, the matrices become integer valued \cite{neumann1992combinatorics,Garoufalidis:2013upa}.\footnote{If the first $(N-1)$ rows of $(A,B)$ matrices are associated to a primitive boundary cycle $\gamma$ which is a non-trivial element of $H_1 (S^3\backslash K, \mathbb{Z}_2)=\mathbb{Z}_2$, then the matrices are integer valued when we choose  $\vec{\mathbb{E}} =\vec{\mathbb{E}}_{\bf m}$. On the  other hand, if the $(A,B)$  are associated to a primitive boundary cycle $\gamma$ which is a trivial element in the $\mathbb{Z}_2$-homology, then the matrices become integer valued when we choose  $\vec{\mathbb{E}} =\vec{\mathbb{E}}_{\bf l}$. Here  the $(A,B)$ are associated to the boundary 1-cycle $\mu$, see eqn~\eqref{boundary holonomy along meridian} and \eqref{Bethe-equations}, which is the generator of the $\mathbb{Z}_2$-homology. } 
\begin{align}
\begin{split}
&\sum_a c_a \mathbb{E}^a_{\bf m}:=- \textrm{diag} \left\{0,c_1, c_1+c_2,\ldots, c_1+\ldots c_{N-1} \right\} + C\; \mathbb{I}_{N\times N}\;,
\\
& C:= \frac{1}N\sum_{a=1}^{N-1} (N-a)c_a\;. \label{basis E}
\end{split}
\end{align}
In the construction of $\mathcal{T}_N[S^3\backslash K, \mu]$, for properly quantized CS levels,  we need to use $(A,B)$ matrices associated to the basis $\{\vec{\mathbb{E}}_{\bf m}\}$.  This is the reason why the basis $\{\vec{\mathbb{E}}_{\bf m}\}$ appears in the  handle-gluing operator in \eqref{factor N-1}. 
On the other hand, the torsion  $ \mathbf{Tor}_{S^3\backslash K} (\tau_{\rm adj},N;\mu)$   can be computed using a state-integral model developed in \cite{Dimofte:2011gm,Dimofte:2012qj} and the result is
\begin{align}
{\bf Tor}_{S^3\backslash K} (\tau_{\rm adj},N;\mu)   = \det \left(A(\vec{\mathbb{E}}^{\bf l})\cdot \Delta_{z''} + B(\vec{\mathbb{E}}^{\bf l}) \cdot \Delta_{z^{-1}}\right) \prod_{}(z_i)^{f''_i} (z''_{i})^{f_i} \label{factor N-2} .
\end{align}
The adjoint torsion on a knot complement, $S^3\backslash K$, also depends on the choice of primitive boundary cycle $\gamma \in H_1 \left(\partial (S^3\backslash K),\mathbb{Z}\right)$ \cite{2015arXiv151100400P}.  We  denote the torsion as    ${\bf Tor}_{S^3\backslash K} (\tau_{\rm adj},N;\gamma)$ to specify the choice. 
The above  torsion is identical to the handling gluing operator in \eqref{factor N-1} except the basis change, from $\vec{\mathbb{E}}_{\bf m}$ to $\vec{\mathbb{E}}^{\bf l}$. The new basis is defined as follows:
\begin{align}
\begin{split}
&\sum_{a=1}^{N-1} c^{(a)} \mathbb{E}^{\bf l}_a= \textrm{diag} \{c^{(1)}, c^{(2)}-c^{(1)},\; \ldots\;,  c^{(N-1)}-c^{(N-2)} , -c^{(N-1)}\}\;. \label{basis E-dual}
\end{split}
\end{align}
Two basis are related by a linear transformation determined by the Cartan matrix $\kappa_{ab}$ of $su(N)$:
\begin{align}
\mathbb{E}^{\bf l}_a  = \sum_{b=1}^{N-1}\kappa_{ab} \mathbb{E}_{\bf m}^b\;, \quad \kappa_{ab}:=\textrm{Tr} (\mathbb{E}^{\bf l}_a\cdot \mathbb{E}^{\bf l}_b) =
\begin{cases}
2, \quad \;\;\; a=b\\
-1, \quad |a-b|=1\\
0, \quad \;\;\; \textrm{otherwise}
\end{cases}
\;. \label{Transformation of two basis}
\end{align}   
Two basis  are conjugate to each other in the following sense
\begin{align}
\textrm{Tr}(\mathbb{E}^a_{\bf m} \cdot \mathbb{E}_b^{\bf l}) = \delta_{b}^a\;.
\end{align} 
The determinant of the Cartan matrix is
\begin{align}
\det_{a,b} (\kappa_{ab})=N\;. \label{Det of kappa}
\end{align}
From equations in \eqref{factor N-1},\eqref{factor N-3}, \eqref{factor N-2},\eqref{Transformation of two basis} and \eqref{Det of kappa}, we finally have 
\begin{align}
\mathcal{H}_{\nu_R = \frac{1}2} (\mathcal{T}_N[S^3\backslash K;\mu])   = N\times {\bf Tor}_{M_3} (\tau_{\rm adj},N;\mu) \;. \label{handle-torsion in knot complement}
\end{align}
\begin{table}[h]
	\begin{center}
		\begin{tabular}{|c|c|}
			\hline 
			$\mathcal{T}_{N}[S^3\backslash  K, \mu]$ on $\mathbb{R}^2\times S^1$& $PSL(N, \mathbb{C})$ CS theory on $S^3\backslash  K$ \\
			\hline \hline
			Bethe equations  & Octahedral gluing equations in \eqref{Bethe-equations}  \\ 
			\hline 
			Bethe vacua & Irreducible flat connections \\
			\hline 
			On-shell twisted superpotential  $\mathcal{W}_{\nu_R =1/2}$& Classical action $S_0$ in \eqref{perturbative expansion of CS ptn}  \\ 
			\hline 
			$S^1$ holonomy $\mathbb{M}$ coupled to flavor $PSU(N)$ & $PSL(N,\mathbb{C})$ holonomy along $\mu$  \\ 
			\hline 
			$\mathbb{L} \simeq \exp \left(\partial_{\log \mathbb{M}} \mathcal{W}_{\nu_R = 1/2}\right)$ in \eqref{longitude holonomy}  & $PSL(N,\mathbb{C})$ holonomy along $\lambda$  \\ 
			\hline 
			$\mathcal{H}_{\nu_R =1/2}$  & $N \times \mathbf{Tor}(\tau_{\rm adj};\mu)$  \\ 
			\hline 
		\end{tabular} \caption {Summary of 3d-3d correspondence for knot complement $S^3\backslash K$. }
		\label{3d-3d for knot complement}
	\end{center}
\end{table}

To arrive at the 3d-3d dictionary in Table~\ref{3d/3d correspondence for twisted index} for handle gluing operator from \eqref{handle-torsion in knot complement}, we only need to show that the way of handling gluing transforms under the gauging operation is equal to the transformation of torsion under the Dehn filling, see \eqref{T[M] under Dehn filling}. 
Depending on whether $\textrm{g.c.d}(P,N) =1$ or not, the hand gluing operator transforms slightly differently under the gauging procedure. Here we only analyze for the simpler case, when $\textrm{g.c.d}(P,N) =1$, and will check that the two transformation rules are identical which prove  the 3d-3d dictionary in Table~\ref{3d/3d correspondence for twisted index}.
On the 3-manifold side, the condition $\textrm{g.c.d}(P,N) =1$ is mapped into the following topological condition 
\begin{align}
H_1 (M_3=(S^3\backslash K)_{P\mu + \lambda}, \mathbb{Z}_N) = \mathbb{Z}_{|\textrm{g.c.d}(P,N)|}  = 0\;.
\end{align}
This is the technical  reason why we assume the topological condition for the simpler 3d-3d relation. It would be an interesting future work to generalize this analysis to derive the 3d-3d relation  for general hyperbolic 3-manifold.

\paragraph{Handle-gluing operator under  gauging} 
Under the $PSU(N)$ gauging with an additional Chern-Simons level $P$ in \eqref{factor N-1},  the handle gluing transform as follows 
\begin{align}
\begin{split}
&\mathcal{H}_{\nu_R = \frac{1}2}(\mathcal{T}_N[M_3= (S^3\backslash K)_{P\mu +\lambda}])  
\\
&=  \mathcal{H}_{\nu_R = \frac{1}2} (\mathcal{T}_N[S^3\backslash K;\mu];\mathbb{M}) \frac{ N \det_{a,b} \partial_{\log m_a} \partial_{\log m_b} \left(\mathcal{W}_{\nu_R = \frac{1}2} [\mathcal{T}_N[S^3\backslash K];\mathbb{M}] + \frac{P}2\textrm{Tr}(\log \mathbb{M})^2\right) } {\prod_{\alpha \in \Lambda^+_{\rm adj}} \left(1-\alpha (\mathbb{M})\right) \left(1-\alpha (\mathbb{M}^{-1}) \right)} \;, 
\\
&\textrm{when g.c.d$(P,N) = 1$} \;.\label{H under PSU(N) gauging}
\end{split}
\end{align} 
The basic structure of the above formula can be understood  from the general localization result in \cite{Closset:2018ghr}. However, there are several subtle points in applying the general result to the $PSU(N)$ gauging case. The subtleties are fixed by requiring to reproduce the known Verlinde formula for pure Chern-Simons theory when we choose $\mathcal{W}_{\nu_R = \frac{1}2}=0$.

Let us explain the expression \eqref{H under PSU(N) gauging} in detail. The matrix  $\mathbb{M}$  is the complexified  holonomy of   the $PSU(N)=SU(N)/\mathbb{Z}_N$  along the $S^1 \subset \mathbb{R}^2 \times S^1$ where $m_{a=1,\ldots, N-1}$ parametrize the holonomy in the following way
\begin{align}
\log \mathbb{M} = \sum_{a=1}^{N-1} \log m_a \mathbb{E}^a_{\bf m} :=\log \textrm{diag}\{u_1(\vec{m}),\ldots, u_N (\vec{m}) \}\;,  \label{log M}
\end{align}
and $\{\mathbb{E}^a_{\bf m}\}_{a=1}^{N-1}$ is a basis given in \eqref{basis E}. Then, 
$\mathcal{W}_{\nu_R = \frac{1}2} [\mathcal{T}_N[S^3\backslash K];\mathbb{M}]$ is the (on-shell) twisted superpotential of $\mathcal{T}_N[S^3\backslash K , \mu]$ in the presence of background  holonomy $\mathbb{M}$ coupled to the $PSU(N)$ symmetry. Following standard notation, $\alpha \in \Lambda^+_{\rm adj}$ denotes positive roots of $psu(N)$
\begin{align}
\prod_{\alpha \in \Lambda^\dagger_{\rm adj}} (1-\alpha (\mathbb{M})) (1-\alpha (\mathbb{M}^{-1})) = \prod_{i\neq j} \left(1-u_i(\vec{m})/u_j(\vec{m})\right)\;.
\end{align}
Bethe-vacua are given by solutions of the following equations modulo a quotient by Weyl group action
\begin{align}
[\mathbb{M}^P  \cdot \mathbb{L}] = [\mathbb{I}_{N\times N}]\;. \label{PSU(N) Dehn filling}
\end{align}
Here $[\mathbb{M}]$ denotes the equivalence class of the $\mathbb{Z}_N$ in $PSU(N) = SU(N)/\mathbb{Z}_N$
\begin{align}
[\mathbb{M}] = [e^{i \theta}\mathbb{M}]\;, \quad \theta =0, \frac{2\pi}N,\ldots, \frac{2\pi (N-1)}{N}\;, \label{PSU(N) Bethe-equations}
\end{align}
and its conjugate $psu(N)$ matrix $\mathbb{L}$ is defined as
\begin{align}
\begin{split}
&\log \mathbb{L} \left(\vec{\ell}(\vec{m})\right) = \sum_{a=1}^{N-1} \log \ell^{(a)} (\vec{m}) \mathbb{E}^{\bf l}_a
\\
& \log \ell^{(a)} := \partial_{\log m_a} \mathcal{W}_{\nu_R = \frac{1}2} [\mathcal{T}_N[S^3\backslash K];\mathbb{M}] \;, \label{longitude holonomy}
\end{split}
\end{align}
where $\{\mathbb{E}^{\bf l}_a \}_{a=1}^{N-1}$ is a basis given in \eqref{basis E-dual}. 
In choosing the Bethe-vacua, we need to choose solutions of \eqref{PSU(N) Bethe-equations} which are not fixed points of the Weyl-action
\begin{align}
\alpha (\mathbb{M}) \neq 0 \quad \textrm{for all $\alpha \in \Lambda^+_{\rm adj}$}\;.
\end{align} 
As a consistency check for the formula, consider a pure $\mathcal{N}=2$ $PSU(N)$ theory with a Chern-Simons level $k >0$. It corresponds to $\mathcal{W}_{\nu_R = \frac{1}2} =0$ and $P=k $ in the formula\footnote{Or one may choose $\mathcal{W}_{\nu_R = \frac{1}2} = \frac{P_1}2 \textrm{Tr} (\log \mathbb{M})^2$ and $P=k-P_1$. The final answer is independent on the choice of $P_1$ since $\mathbb{L}(\vec{l})  = \mathbb{M}(\vec{m})^{P_1}$ if $\log \ell^{(a)} = \partial_{\log m_a} \frac{P_1}2 \textrm{Tr} (\log \mathbb{M})^2$. This gives a zero-th order consistency check for the formulae in 
	\eqref{H under PSU(N) gauging} and	\eqref{PSU(N) Dehn filling}.}:
\begin{align}
&\mathcal{H}(PSU(N)_k) = \frac{  N \det_{a,b} \partial_{\log m_a} \partial_{\log m_b} ( \frac{k}2 \sum_{i=1}^N(\log u_i)^2) } {\prod_{i\neq j} (1-e^{u_i}/e^{u_j})} = \frac{k^{N-1}}{ \prod_{i\neq j} (1-e^{u_i}/e^{u_j})}\;,
\\
&\mathcal{Z}^{\nu_R = \frac{1}2}_{p=0, g} (PSU(N)_k ) =  \sum_{[\mathbb{M}^k]=[\mathbb{I}_{N\times N}];u_{i+1}>u_i }\frac{k^{(g-1)(N-1)}}{ \prod_{i\neq j} (1-e^{u_i}/e^{u_j})^{g-1}}
\end{align}
Note that the summation is over $\{m_a\}_{a=1}^{N-1}$ satisfying the above constraints modulo $\mathbb{M} \sim e^{\frac{2\pi i n}{N}}\mathbb{M}$ with $n=0,\ldots, N-1$. Recall the definition of $u_i (\vec{m})$ in \eqref{log M}.  For the case when g.c.d$(k,N) = 1$, the expression is actually equivalent to the following  Verlinde formula \cite{blanchet2005spin}:
\begin{align}
\begin{split}
&\mathcal{Z}^{\nu_R = \frac{1}2}_{p=0, g} (PSU(N)_k )  = \frac{(\textrm{g.c.d}(k,N))^g}{N^g} \mathcal{Z}^{\nu_R = \frac{1}2}_{p=0, g} (SU(N)_k ) \;,
\\
&\mathcal{Z}^{\nu_R = \frac{1}2}_{p=0, g} (SU(N)_k ) =
\begin{cases}
\left( \frac{k}2 \right)^{g-1} \sum_{j=1}^{k+1} |\sin \frac{j \pi} k|^{2-2g} \;, \quad N=2\\
\left( \frac{N}k \right)^{g} \sum_{S \subset\{1,\ldots, k\};|S|=N} \prod_{s\in S,t \in S^c}\big{|} 2 \sin \pi \frac{s-t}{k}\big{|}^{g-1}\;, \quad N\geq 3
\end{cases}
\end{split}
\end{align}
In the expression, we take into account of the 1-loop CS level shift $k_{\mathcal{N}=0}+N = k_{\mathcal{N}=2}$ comming  from integrating out the auxiliary massive gaugino in the $\mathcal{N}=2$ vector multiplet. 
\begin{table}[h]
	\begin{center}
		\begin{tabular}{|c|c|}
			\hline 
			$\mathcal{T}_{N}[(S^3\backslash  K)_{P \mu+\lambda}]$ on $\mathbb{R}^2 \times S^1$ & $PSL(N, \mathbb{C})$ CS theory on $(S^3\backslash  K)_{P\mu+\lambda}$ \\
			\hline \hline
			Gauging $PSU(N)_P$ of  $\mathcal{T}_{N}[S^3\backslash  K;\mu]$ & Dehn filling on $S^3\backslash K$ with slope $P$  \\ 
			\hline 
			Bethe equations in \eqref{PSU(N) Dehn filling} & Gluing equations  in \eqref{gluing-equations-N=2-2}\\
			\hline 
			Bethe vacua & Irreducible flat connections \\
			\hline 
			$\mathcal{H}_{\nu_R =1/2}$ under the gauging in \eqref{H under PSU(N) gauging}  & $\mathbf{Tor}(\tau_{\rm adj})$ unde the Dehn filling in \eqref{torsion under Dehn filling}  \\ 
			\hline 
		\end{tabular} \caption {Gauging/Dehn filling in 3d-3d correspondence. Here we assume $\textrm{g.c.d}(P,N)=1$. }
		\label{3d-3d for Dehn filled manifold}
	\end{center}
\end{table}
\paragraph{Adjoint torsion under  Dehn filling}  The transformation rule \eqref{H under PSU(N) gauging} of the handle gluing operator is  exactly the  same as the way the adjoint torsion  transforms under the Dehn filling along a slope $P\mu +\lambda$:
\begin{align}
\begin{split}
&{\bf Tor}_{M = (S^3 \backslash K)_{P\mu + \lambda}} (\tau_{\rm adj},N)  
\\
&=  {\bf Tor}_{S^3\backslash K} (\tau_{\rm adj},N;\mu)  \frac{ N \det_{a,b} \partial_{\log m_a} \partial_{\log m_b} \left(S_0(S^3\backslash K ;\mathbb{M} ) + \frac{P}2 \textrm{Tr}(\log \mathbb{M})^2 \right)} {\prod_{i\neq j} (1-e^{u_i}/e^{u_j})} \;. \label{torsion under Dehn filling}
\end{split}
\end{align} 
On the 3-manifold side,  $S_0 (S^3\backslash K, \mathbb{M})$ is the classical Chern-Simons action for a flat-connection on $S^3\backslash K$ with boundary $PSL(N,\mathbb{C})$ holonomy $\mathbb{M}$. According to a 3d-3d relation \cite{Dimofte:2010tz}, the action is identical to the on-shell twisted superpotential:
\begin{align}
S_0 (S^3\backslash K ;\mathbb{M}) = \mathcal{W}_{\nu_R = \frac{1}2} [\mathcal{T}_N[S^3\backslash K];\mathbb{M}] \;.
\end{align}
The matrix $\mathbb{L}$ in \eqref{longitude holonomy} corresponds to the $PSL(N,\mathbb{C})$ holonomy of the flat connection along the  dual  boundary 1-cycle $\gamma_{\rm dual}$, which is chosen as longitude $\lambda$. Then, the Bethe equations in \eqref{PSU(N) Dehn filling} corresponds to the conditions that  a $PSL(N,\mathbb{C})$ flat connection on $S^3\backslash K$ can be  extended to the Dehn filled closed manifold, see  \eqref{gluing-equations-N=2-2} in appendix \ref{app : torsion computation}. So  the solutions to the Bethe equations give flat connections on the closed 3-manifold. 
The numerator in \eqref{torsion under Dehn filling} comes from the change of boundary 1-cycle. Under the  change of boundary 1-cycle, in general,  the adjoint torsion transforms as follows   \cite{porti1997torsion}:
\begin{align}
\begin{split}
{\bf Tor}_{S^3\backslash K} (\tau_{\rm adj} , N;P\mu +Q \lambda) &= {\bf Tor}_{S^3\backslash K} (\tau_{\rm adj},N;\mu)  \times \det_{a,b}\left(\frac{\partial (P \log m_a +Q \sum_c \kappa_{ac}\log \ell^{(c)})}{\partial \log m_b}\right) \;,
\\
&= {\bf Tor}_{S^3\backslash K} (\tau_{\rm adj},N;\mu)  \det \left(P \mathbb{I} +Q \kappa \cdot \frac{\partial \log \ell}{\partial \log m}\right) \;. \label{Torsion under 1-cycle change}
\end{split}
\end{align} 
The basis $\vec{\mathbb{E}}^{\bf l}$ is given in \eqref{basis E-dual} and  $\kappa$ is the Cartan matrix of $su(N)$ as defined in \eqref{Transformation of two basis}.  Using the following facts
\begin{align}
\det \kappa = N, \quad  (\kappa^{-1})^{ab} = \partial_{\log m_a} \partial_{\log m_b} \left(\frac{1}2 \Tr (\log \mathbb{M})^2\right) \; \;\textrm{and}\; \;\log \ell^{(a)} = \partial_{\log m_a} S_0 [S^3\backslash K, \mathbb{M}], \nonumber
\end{align}
we verify that
\begin{align}
\det \left(P\mathbb{I} + \kappa \cdot \frac{\partial \log \ell}{\partial \log m}\right) = N \times \det_{a,b} \partial_{\log m_a} \partial_{\log m_b} \left(\frac{P}2 \Tr (\log\mathbb{M})^2 + S_0 [S^3\backslash K, \mathbb{M}]\right)\;.
\end{align}
The denominator  in \eqref{torsion under Dehn filling} comes  from the  effect of Dehn filling \cite{2015arXiv151100400P}:
\begin{align}
{\bf Tor}_{M_3= (S^3 \backslash K)_{P\mu + \lambda}} (\tau_{\rm adj},N) = \frac{{\bf Tor}_{S^3\backslash K} (\tau_{\rm adj} , N;P\mu + \lambda)}{\prod_{i\neq j} (1-e^{u_i}/e^{u_j})}\;. \label{Torsion under Dehn filling}
\end{align}
Combining \eqref{handle-torsion in knot complement},\eqref{H under PSU(N) gauging} and \eqref{torsion under Dehn filling}, we finally derive the 3d-3d relation the for handle-gluing operator in Table \ref{3d/3d correspondence for twisted index}.


%

%
%
%

%
%
%
%
%
\subsection{Consistency check: Integrality of twisted indices}
Generally, the partition function $\mathcal{Z}^{\nu_R}_{p,g} $ for $p=0$ should be integer-valued since it counts the number of ground states of 3d SCFT on $\Sigma_g$ with signs.  
In  the expression \eqref{dmicro-sum-over-flat-connetions}, the integrality of  $\mathcal{Z}^{\nu_R = \frac{1}2}_{p=0,g} (\mathcal{T}_N[M_3])$ is far from obvious.  We check the integrality for several examples below, and naturally we conjecture it is always true. This is a curiosity, and the integral property of torsion has been already reported in the mathematical literature \cite{kitano2016some}. One crucial difference is that they consider torsions in the fundamental  representation, while  we consider here the  adjoint representation. 

\paragraph{Example of  $M_3= (S^3\backslash \mathbf{4}_1)_{5\mu + \lambda}$ and $N=2$:} 
The corresponding 3d gauge theory  was proposed in \cite{Gang:2017lsr}\footnote{In this example, the 3d theory has an additional $U(1)_{\rm top}$ flavor symmetry whose coserved charge counts monopole charge of the $U(1)$ gauge field. Such an accidental  bonus symmetry can appear in $\mathcal{T}_N[M_3]$ theory only for small $N$ as  argued in \cite{Gang:2018wek}. In this example, the IR superconformal R-symmetry $U(1)_R$ is a mixture of compact $SO(2)$  R-symmetry originated from 6d $SO(5)$ R-symmetry and the accidental $U(1)_{\rm top}$. The IR R-symmetry charge is not properly quantized, hence we can not use it for the topological twisting along $\Sigma_g$. In this example, we use the $SO(2)$ symmetry for the topological twisting. The 3d-3d relation Table~ \ref{3d/3d correspondence for twisted index} works for the twisted index using the $SO(2)$ R-symmetry which is always identical to the IR superconformal R-symmetry for sufficiently large $N$.}
\begin{align}
\begin{split}
&\mathcal{T}_{N=2}[ (S^3\backslash \mathbf{4}_1)_{5\mu + \lambda}] 
\\
&= (\textrm{$\mathcal{N}=2$ $U(1)$ vector coupled to a chiral $\Phi$ of charge $+1$ with CS level $k= - 7/2$})\;. \label{T-thurston}
\end{split}
\end{align}
The Witten index for the theory is \cite{Intriligator2013}
\begin{align}
|k|+1/2 = 4\;.
\end{align}
Therefore,  there are four Bethe vacua of the theory which are given as solutions to the  following algebraic equation extremizing the twisted superpotential  ($z:=e^{Z}$)
\begin{align}
\exp \big{(}\partial_Z \mathcal{W}_{\nu_R = \frac{1}2} (Z)\big{)} = \frac{1-z}{z^4}=1\;, \quad \mathcal{W}_{\nu_R = \frac{1}2} (Z) = \textrm{Li}_2 (e^{-Z}) - \frac{3}2 Z^2 + i \pi Z . 
\end{align}
The four solutions are
\begin{align}
\{ \hat{z}_{\alpha} \}_{\alpha=1}^4 = \{ 0.248126 - 1.03398 i,\; 0.248126 - 1.03398 i ,\; -1.22074,\;0.724492 \}.
\end{align}
The handle gluing operator is \cite{Closset:2017zgf,Closset:2018ghr}
\begin{align}
\mathcal{H}^{(\alpha)}_{\nu_R = \frac{1}2} = \frac{(1-\frac{1}z)}z  \left(\partial_Z \partial_Z \mathcal{W}_{\nu_R = \frac{1}2} (Z) \right)  \bigg{|}_{z= \hat{z}_{\alpha}}= \frac{4-3z}{z^2} \bigg{|}_{z= \hat{z}_{\alpha}}\;.
\end{align}
Their  numerical values are
\begin{align}
\begin{split}
&\{ \mathcal{H}^{(\alpha)}_{\nu_R = \frac{1}2} \textrm{ for } \mathcal{T}_{N=2}[M_3= (S^3\backslash \mathbf{4}_1)_{5\mu + \lambda}] \}_{\alpha=1}^4 
\\
&= \{ -3.81076 - 1.13799 i, \; -3.81076 + 1.13799 i,\; 
5.14169,\; 3.47983 \}\;. \label{torsion-41-51}
\end{split}
\end{align}
Comparing the analytic  torsions in \eqref{Torsion-thurston-N=2} for four irreducible $PSL(2,\mathbb{C})$ flat-connections on $M_3$, we confirm the 3d-3d relation for the handle-gluing operator in Table~\ref{3d/3d correspondence for twisted index}.
Applying these results to \eqref{dmicro-sum-over-flat-connetions}, we have
\begin{align}
\begin{split}
&\{ \mathcal{Z}_{p=0,g}^{\nu_R =\frac{1}2} (\mathcal{T}_{N=2}[M_3 = (S^3 \backslash \mathbf{4_1})_{5\mu +\lambda}]) \}_{g=0}^\infty
\\
&=  \big{\{} 0_{g=0},\; 4_{g=1}, \;1_{g=2} , \; 65_{g=3}, \; 97_{g=4},\; 1045_{g=5},\;\ldots  \big{\}} \;.
\end{split}
\end{align}
Note that these are all  integers!  Using the explicit formulae in  Appendix \ref{app : torsion computation}, one can compute  ${\bf Tor}^{(\alpha)}_{M_3= (S^3\backslash \mathbf{4}_1)_{P\mu +Q\lambda}}[\tau_{\rm adj},N=2]$ for arbitrary $(P,Q)$s  and check that the  $\mathcal{Z}^{\nu_R = \frac{1}2}_{p=0,g} (\mathcal{T}_N[M_3])$ in \eqref{dmicro-sum-over-flat-connetions} is always integer when the $P$ is odd. The oddness of $P$ is equivalent to the topological condition of vanishing $H_1 \left(M_3= (S^3\backslash \mathbf{4}_1)_{P\mu +Q\lambda},\mathbb{Z}_2\right)$. This provides a non-trivial consistency check for the 3d-3d relation for handle-gluing operator in Table~\ref{3d/3d correspondence for twisted index}.
\section{Full perturbative $1/N$ expansion of twisted partition functions}\label{Sec:LargeN}
Combining the expression  in \eqref{dmicro-sum-over-flat-connetions} with a mathematical result on asymptotic properties of the analytic torsion, we  determine the full perturbative $1/N$ corrections to the twisted partition functions. The final perturbative expression is given in  \eqref{final large N for p>0}  and \eqref{final large N for p=0}.
\subsection{Two dominant Bethe-vacua from the hyperbolic structure}
In $\mathcal{T}_N[M_3]$ theory for a hyperbolic $M_3$, there are two special Bethe-vacua which correspond  to two irreducible flat-connections, $\mathcal{A}^{\rm geom}_N$ and $\mathcal{A}^{\overline{\rm geom}}_N$, on $M_3$. The flat connections can be constructed using the unique hyperbolic structure on $M_3$
\begin{align}
\mathcal{A}^{\rm geom}_N = \tau_{N} \cdot (\omega + i e)\;, \quad  \mathcal{A}^{\rm \overline{geom}}_N = \tau_{N} \cdot (\omega - i e)\;, \label{Two irreducible flat connections}
\end{align}
$\omega$ and $e$ are spin-connections and vielbein of the unique hyperbolic metric on $M_3$. They form two $PSL(2,\mathbb{C})$ irreducible flat-connections $\omega \pm i e$, which are lifted to two $PSL(N,\mathbb{C})$ irreducible flat connections,  $\mathcal{A}^{\rm geom}_N$ and $\mathcal{A}^{\rm \overline{geom}}_N$, via the $N$-dimensional irreducible representation $\tau_N$ of $su(2)$. We define
\begin{align}
\tau_m := \textrm{Sym}^{\otimes (m-1)} \tau_2\;, \quad \tau_2 := \textrm{fundamental representation of $su(2)$}\;. \label{tau-m}
\end{align}
The flat connection $\mathcal{A}^{\rm geom}_{N=2}$ is actually identical to the flat connection $\rho_{\rm geom}$ given in \eqref{discrete-faithful rep}. 
 The fibering  operators for the Bethe-vacua are
\begin{align}
\begin{split}
&|\mathcal{F}^{\rm geom}_{\nu_R = \frac{1}2} (\mathcal{T}_N[M_3])| =\bigg{|}\exp \bigg{(} \frac{CS[\mathcal{A}_N^{\rm geom}]}{4\pi i }\bigg{)}  \bigg{|}=  \exp \bigg{(} - \frac{N^3-N}{12 \pi} \textrm{vol}(M_3)\bigg{)} \;, \quad
\\
&|\mathcal{F}^{\overline{\textrm{geom}}}_{\nu_R = \frac{1}2} (\mathcal{T}_N[M_3])| =  \bigg{|}\exp \bigg{(} \frac{CS[\mathcal{A}_N^{\overline{\textrm{geom}}}]}{4\pi i }\bigg{)}  \bigg{|}=\exp \bigg{(}  \frac{N^3-N}{12 \pi} \textrm{vol}(M_3)\bigg{)}\;. \label{Fibering-for-two-Vacua}
\end{split}
\end{align}
Refer to \cite{Gang:2014qla} for the computation of Chern-Simons functionals for these  two flat connections. Moreover, it is known that 
\begin{align}
\bigg{|}\exp \bigg{(} \frac{CS[\mathcal{A}_N^{\rm geom}]}{4\pi i }\bigg{)}  \bigg{|} <\bigg{|}\exp \bigg{(} \frac{CS[\mathcal{A}_N^{\alpha}]}{4\pi i }\bigg{)}  \bigg{|}<\bigg{|}\exp \bigg{(} \frac{CS[\mathcal{A}_N^{\overline{\rm geom}}]}{4\pi i }\bigg{)}  \bigg{|} \label{property of two connections}
\end{align}
for arbitrary $PSL(N,\mathbb{C})$ flat connection $\mathcal{A}_N^\alpha$ other than the two special irreducible flat connections.
\paragraph{Large $N$ expansion of $|\mathcal{H}^{\textrm{geom}}| = |\mathcal{H}^{\overline{\textrm{geom}}}|$:}
The two flat-connections are simply related by complex conjugation and so are their analytic torsions
\begin{align}
\begin{split}
&{\bf Tor}^{(\rm geom)}_{M_3} (\tau_{\rm adj},N) = e^{i \theta_{N,M_3}}|{\bf Tor}^{(\rm geom)}_{M_3} (\tau_{\rm adj},N)|\;, \quad  
\\
&{\bf Tor}^{(\overline{\textrm{geom}})}_{M_3} (\tau_{\rm adj},N) = e^{-i \theta_{N,M_3}}|{\bf Tor}^{(\overline{\textrm{geom}})}_{M_3}(\tau_{\rm adj},N)|\;, 
\label{relation-between-two-torsions}
\end{split}
\end{align}
where $e^{i \theta_{N,M_3}}$ is a phase factor. 
From the  branching rule
\begin{align}
\left(\textrm{$\tau_{\rm adj}$  of $su(N)$}\right) = \oplus_{m=1}^{N-1} (\tau_{2m+1} \textrm{ of $su(2)$})\;, \label{branching rule}
\end{align}
we can decompose the analytic torsion for $PSL(N, \mathbb{C})$ into products of analytic torsions for $PSL(N=2, \mathbb{C})$
\begin{align}
\log |{\bf Tor}^{(\rm geom)}_{M_3} (\tau_{\rm adj},N)|= \sum_{m=1}^{N-1} \log |{\bf Tor}^{(\rm geom)}_{M_3} (\tau_{2m+1},N=2)|\;.
\end{align}
The large $m$ expansion of the torsion  $|{\bf Tor}^{(\rm geom)}_{M_3} (\tau_{2m+1},N=2)|$  were studied in  \cite{muller2012asymptotics},
\begin{align}
\begin{split}
&\log |{\bf Tor}^{(\rm geom)}_{M_3}(\tau_{2m+1},N=2)| \;,  \qquad  (\textrm{for $m \geq 1$})
\\
&=   \frac{1}{\pi} \textrm{vol}(M_3) (m^2+m)+\log |{\bf Tor}_{M_3}(N=1)| -\sum_{[\gamma]}\sum_{k=1}^m \log  |1- e^{-k \ell_{\mathbb{C}}(\gamma)}|\;.
\label{N=2 torsion asymptotic}
\end{split}
\end{align}
Here $\textrm{\bf Tor}_{M_3} (N=1)$ is the scalar torsion, torsion associated to the trivial bundle, on $M_3$. The large $m$ expansion can be numerically checked  up to $o(m)$  terms for many examples of $M_3= (S^3\backslash \mathbf{4}_1)_{P\mu + Q\lambda}$ using the explicit expression  given in Appendix \ref{app : torsion computation}.
Combining the branching rule \eqref{branching rule} and the asymptotic expansion, we have the following large $N$ expansion of the adjoint torsion
\begin{align}
\begin{split}
&\log |{\bf Tor}^{(\rm geom)}_{M_3} (\tau_{\rm adj},N)|
\\
&=\sum_{m=1}^{N-1} \bigg{(}  \frac{1}{\pi} \textrm{vol}(M) (m^2+m)+\log |{\bf Tor}_{M_3}(N=1)| -\sum_{[\gamma]}\sum_{k=1}^m \log  |1- e^{-k \ell_{\mathbb{C}}(\gamma)}| \bigg{)}\;,
\\
&=  \frac{1}\pi \textrm{vol}(M_3)(N^3-N) +(N-1)\log |{\bf Tor}_{M_3}(N=1)|  + \mathfrak{Re}\sum_{[\gamma]}\sum_{s=1}^\infty \sum_{k=1}^{N-1} \frac{(N-k)}s  e^{-s k \ell_{\mathbb{C}}(\gamma) }\;,
\\
&= \frac{\textrm{vol}(M_3)}{3\pi} (N^3-N) - a(M_3) (N-1)-b(M_3)+c(M_3;N)\;.  \label{large-N-of-torsion}
\end{split}
\end{align}
Here we have defined
\begin{align}
\begin{split}
& a(M_3) := a_1(M_3)+a_{2}(M_3)  \quad \textrm{where}
\\
& \qquad \qquad  a_1(M_3) :=  -\log |{\bf Tor}_{M_3}(N=1)|  \;, \quad a_2(M_3):=\sum_{[\gamma]} \sum_{m=1}^\infty \log  |1-e^{-m \ell_{\mathbb{C}}(\gamma)}|\;,
\\
&b(M_3):= \mathfrak{Re}  \sum_{[\gamma]} \sum_{s=1} \frac{1}{s}  \left(\frac{ e^{- s \ell_{\mathbb{C}}(\gamma)} }{1-e^{-s \ell_{\mathbb{C}} (\gamma)}} \right)^2\;,
\\
&c(M_3;N) := \mathfrak{Re}  \sum_{[\gamma]} \sum_{s=1} \frac{1}{s} \bigg{(}\frac{  e^{-\frac{s(N+1)}2   \ell_{\mathbb{C}}(\gamma)}}{1-e^{-s \ell_{\mathbb{C}}(\gamma)}} \bigg{)}^2\;. \label{perturbative coefficients (a,b,c)}
\end{split}
\end{align}
Note that $c(M_3;N)$ is exponentially suppressed at large $N$.  In the formulae above, $[\gamma]$ runs over the nontrivial primitive conjugacy classes of $\pi_1 (M_3)$. The $PSL(2,\mathbb{C})$ flat connection $\mathcal{A}^{\rm geom}_{N=2}$ on $M_3$ gives a homomorphism $\rho_{\rm geom}$
\begin{align}
\rho_{\rm geom}  \in \textrm{Hom}[\pi_1 (M_3) \rightarrow PSL(2,\mathbb{C})]\;.
\end{align}
The complex length $\ell_{\mathbb{C}}$ of $\gamma$ is defined by
\begin{align}
\textrm{Tr} \rho_{\rm geom} ( \gamma) = 2  \cosh \left(\frac{1}2 \ell_{\mathbb{C}} (\gamma) \right) \;, \quad \mathfrak{Re}\ell_{\mathbb{C}} > 0\;.
\end{align}
 In the above expressions, we assume the convergence of the infinite summation over $[\gamma]$. If it does not converge, we need to use the following formula instead of \eqref{N=2 torsion asymptotic}
\begin{align}
\begin{split}
&\log |{\bf Tor}_{M_3}^{(\textrm{geom})}(\tau_{2m+1},N=2)| \;,  \qquad  (\textrm{for $m \geq 2$})
\\
&=   \frac{1}{\pi} \textrm{vol}(M_3) (m^2+m-6)+ \log |{\bf Tor}_{M_3}^{(\textrm{geom})}(\tau_{5},N=2)| -\sum_{[\gamma]}\sum_{k=3}^m \log  |1- e^{-k \ell_{\mathbb{C}}(\gamma)}|\;.
\end{split}
\end{align}
The infinite sum here is proven to be always absolutely convergent \cite{muller2012asymptotics}. If one wants to use this safer infinite sum, we only need to replace $\log |{\bf Tor}_{M_3}(N=1)|$ in  the above expressions in the following way:
\begin{align}
\begin{split}
&\log |{\bf Tor}_{M_3} ( N=1)|
\\
&\longrightarrow  \log |{\bf Tor}_{M_3}^{(\textrm{geom})}(\tau_5, N=2)| +\sum_{[\gamma]}\sum_{k=1}^2 \log  |1- e^{-k \ell_{\mathbb{C}}(\gamma)}|-\frac{6}\pi \textrm{vol}(M_3) \;.
\end{split}
\end{align}
A superiority of the large $N$ expansion formulae in \eqref{large-N-of-torsion}  is the appearance of more familiar group  theoretical factors, $(N^3-N)$ and $(N-1)$, which may provide  important clues for understanding the M-theoretical  origin of each term in the expansion.

\subsection{For $p>0$} 
From \eqref{dmicro-sum-over-flat-connetions} and \eqref{property of two connections}, we have following large $N$ expansion of the twisted partition function with $p \in 2\mathbb{Z}_{>0}$ :
\begin{align}
\begin{split}
&\mathcal{Z}^{\nu_R = \frac{1}2}_{g,p \in 2 \mathbb{Z}_{> 0}} (\mathcal{T}_{N}[M_3])=  \sum_{ \mathcal{A}^\alpha \in \chi^{\rm irred} (N,M_3)} N^{g-1}  \exp \left(p \frac{CS[\mathcal{A}^\alpha]}{4\pi i } \right)  {\bf Tor}_{M_3}^{(\alpha)}  (\tau_{\rm adj},N)^{g-1}\;.
\\
& = N^{g-1} \exp \left( p \frac{CS[\mathcal{A}^{\overline{\rm geom}}_N;M_3]}{4\pi i } \right )  {\bf Tor}_{M_3}^{(\overline{\textrm{geom}})} (\tau_{\rm adj},N)^{g-1}
\\
& \quad \;\;+(\textrm{exponentially smaller corrections when $N \rightarrow \infty $}) 
\end{split}
\end{align}
From \eqref{Fibering-for-two-Vacua} and \eqref{large-N-of-torsion}, we obtain the following full perturbative $1/N$ expansion 
\begin{equation}
\boxed{
\begin{array}{rcl}
&&\big{|}\mathcal{Z}^{\nu_R = \frac{1}2}_{g,p \in 2 \mathbb{Z}_{\geq 1}} (\mathcal{T}_{N}[M_3]) \big{|}  
\\
&&\xrightarrow{ \quad N\rightarrow \infty \quad }  \exp \bigg{(} \frac{\big{(}4(g-1)+ p \big{)}\textrm{vol}(M_3) }{12\pi} (N^3-N) -(g-1)a(M_3) (N-1)
\\
&&\qquad \qquad \qquad  \qquad  -(g-1)b(M_3) +(g-1) \log N +(g-1) c(M_3;N)\bigg{)}  \times  \bigg{(}1+  e^{- \big{(}\ldots \big{)}} \bigg{)}\;.
\end{array}
}
\label{final large N for p>0}
\end{equation}
This expression is valid for any closed hyperbolic 3-manifold $M_3$ with trivial $H_1 (M,\mathbb{Z}_N)$. Here the perturbative expansion coefficients $(a,b,c)$ are given in \eqref{perturbative coefficients (a,b,c)}. These coefficients are determined by the 
complex length spectrum $\{\ell_{\mathbb{C}} (\gamma) \}$ on the hyperbolic 3-manifold.  We denote exponentially suppressed terms at large $N$ terms  by 
$e^{- (\ldots)}$ as above. Note that the leading term nicely reproduces the gravity free energy  \eqref{leading-I-bolt} for AdS-Taub-Bolt$_+$ solution. 
Two remarkable properties of the above asymptotic expansion are worth highlighting: 

1. The perturbative expanson in $1/N$ terminates at finite order $o(N^0)$. 

2. Logarithmic correction to the $\log \mathcal{Z}$  is $(g-1)\log N$.

\subsection{For $p=0$ and $g>1$}
For $p=0$  and $g>1$ case, we expect that only two irreducible flat-connections, $\mathcal{A}_N^{\rm geom}$ and $\mathcal{A}_N^{\overline{\rm geom}}$, equally give the most dominant contributions to the twisted index at large $N$:
\begin{align}
\begin{split}
&\mathcal{Z}^{\nu_R = \frac{1}2}_{g,p=0} (\mathcal{T}_{N}[M_3])=  \sum_{ \mathcal{A}^\alpha \in \chi^{\rm irred} (N,M_3)} N^{g-1}\left( {\bf Tor}_{M_3}^{(\alpha)}  (\tau_{\rm adj},N)\right)^{g-1}\;.
\\
& = N^{g-1} \big{[}({\bf Tor}^{(\rm geom)}_{M_3}(\tau_{\rm adj},N))^{1-g} +({\bf Tor}^{(\overline{\rm geom})}_{M_3}( \tau_{\rm adj},N) )^{g-1} \big{]}
\\
& \quad \;\;+(\textrm{exponentially smaller corrections when $N \rightarrow \infty $})\;.
\end{split}
\end{align}
From \eqref{relation-between-two-torsions} and \eqref{large-N-of-torsion}, we obtain the following simple large $N$ asymptotic expansion 
\begin{equation}
\boxed{
\begin{array}{rcl}
&&|\mathcal{Z}^{\nu_R = \frac{1}2}_{g,p=0} (\mathcal{T}_{N}[M_3])| 
\\
&&\xrightarrow{ \quad N\rightarrow \infty \quad }   2 \cos \big{(}(1-g)\theta_{N,M_3} \big{)} 
\\
&&\qquad \qquad   \times  \exp \bigg{(} (g-1) \big{(}\frac{\textrm{vol}(M_3) }{3\pi} (N^3-N)  -a(M_3) (N-1)-b(M_3)+ \log N - c(M_3;N) \big{)}\bigg{)}
\\
&& \qquad \qquad   \times  \bigg{(}1+ e^{- \big{(}\ldots \big{)}} \bigg{)}\;. \label{final large N for p=0}
\end{array}
}
\end{equation} 
Again, this expression is valid for any closed hyperbolic 3-manifold $M_3$ with trivial $H_1 (M,\mathbb{Z}_N)$.  The perturbative expansion coefficients $(a,b,c)$ are given in \eqref{perturbative coefficients (a,b,c)}. Note that the leading term nicely reproduce the gravity free-energy \eqref{leading-I-bolt} for AdS-Taub-Bolt solution when $p=0$ or equivalently the Bekenstein-Hawking entropy \eqref{large-N-of-torsion} for  magnetically charged AdS blackhole \cite{Gang:2018hjd}.
Two remarkable properties of the above asymptotic expansion are worth singling out:

1. Modulo an overall  factor $2 \cos \big{(}(1-g)\theta_N \big{)}$, the $1/N$  expanson terminates at  $o(N^0)$. 

2. Logarithmic correction to the $\log \mathcal{Z}$ is $(g-1)\log N$.
\\
The logarithmic correct will be reproduced from a supergravity analysis.,

\section{Logarithmic corrections from Supergravity}\label{Sec:Oneloop}

The Bekenstein-Hawking entropy of any black hole is proportional to the area of its event horizon.  This term, however universal, should be viewed as the leading contribution in a quantum expansion.  Studying corrections to the Bekenstein-Hawking entropy is, therefore,  crucial for a quantum understanding of black holes and for clarifying the microscopic degrees of freedom responsible for the
macroscopic entropy.  Within all the corrections that might be present, logarithmic corrections are particularly central because they are  determined by the massless degrees of freedom of the gravitational theory and are fairly independent of the details of its  ultraviolet completion. In the context of asymptotically flat black holes the computations of logarithmic corrections to the black hole entropy have convincingly provided an infrared window into ultraviolet physics; in every case the supergravity (IR) results have perfectly matched the string theory prediction (see, for example, \cite{Banerjee:2010qc,Banerjee:2011jp,Sen:2011ba,Sen:2012cj,Sen:2012dw} and references therein). Given the recent advances in our understanding of  the microscopic description of certain asymptotically AdS black hole entropy via field theory localization, it is of paramount importance that we extend those remarkable results for asymptotically flat black holes to the context of asymptotically $AdS$ black holes. Doing so will advance the inherent promise of the AdS/CFT correspondence of providing a non-perturbative path to quantum gravity in asymptotically AdS spacetimes. Indeed, for a class of black holes some progress has been reported in \cite{Liu:2017vbl} after preliminary explorations in  \cite{Liu:2017vll,Jeon:2017aif}. 

Let us also point out that everything we stated about black holes above applies to other supergravity backgrounds where we consider instead of the entropy the on-shell action and one-loop quantum corrections around the solution. Indeed, an early application matched the logarithmic in $N$ term in the free energy of  a large class of 3d Chern-Simons matter theory with the one-loop eleven dimensional supergravity computation \cite{Bhattacharyya:2012ye}. Similarly, the computations we perform apply not only to the extremal, magnetically charged asymptotically AdS${}_4$ black hole reviewed in section \ref{Sec:holography} but also to the Taub-Bolt-AdS${}_4$ solution when embedded in  eleven dimensional supergravity. We have already matched the leading part of the on-shell action in section \ref{Sec:holography}  (see also  \cite{Gang:2018hjd}) and in what follows we will match the coefficient of the logarithmic in $N$ term as computed from the one-loop effective action to the microscopic answer following from the appropriate partition function computed in section \ref{Sec:LargeN}. 

Given the wide range of diverse topics covered in this manuscript we provide a brief  review of the main arguments involved in computations of logarithmic corrections to black hole entropy in the context of  11d supergravity, we refer the reader to some relevant work including  \cite{Liu:2017vll,Liu:2017vbl,Jeon:2017aif,Bhattacharyya:2012ye} for more details.

In this section we first make a general comment about the nature of logarithmic terms in one-loop effective actions. We highlight  that in odd dimensional spaces  only zero modes  and boundary terms can contribute to the logarithmic expression. We make the assumption that the whole contribution to the one-loop effective action comes from the asymptotic $AdS_4$ region as was the case  in \cite{Bhattacharyya:2012ye} for the $AdS_4$ solution and in \cite{Liu:2017vbl} for the magnetically charged asymptotically $AdS_4$ black hole case. This assumption will turn out {\it a posteriori} to lead to the  answer which agrees with the field theory expectation.  It does, however, deviates from the standard paradigm where logarithmic corrections are computed using exclusively the near horizon geometry; we believe that this is a feature of asymptotically AdS black holes that deserves further scrutiny.

\subsection{Robustness of logarithmic terms in one-loop effective actions}
To construct the one-loop effective action we integrate, in the path integral, the quadratic fluctuations around the black hole supergravity background. This process leads to the computation of determinants of the corresponding operators. For a given kinetic operator $\CO$ one naturally defines  the logarithm of its  determinant  as
\bea
\frac{1}{2} \: \ln \text{det}' \CO = \frac{1}{2} \sum_n {}' \: \ln \kappa_n
\eea 
where prime denotes that the sum is over non-vanishing eigenvalues, $\kappa_n$,  of $\CO$. It is convenient to define the heat Kernel of the operator $\CO$ formally  as 
\bea
K (\tau) = e^{- \tau \CO} = \sum_n \: e^{- \kappa_n \tau} \mid \phi_n \rangle \langle \phi_n \mid. 
\eea

As emphasized already more than three decades ago in an exquisitely pedagogical manner by Duff and Toms in  \cite{Duff:1982gj}, the heat kernel contains information on both the non-zero modes as well as the zero modes. There  is a very clear prescription widely utilized by Sen and collaborators (see for example, \cite{Sen:2011ba,Sen:2012cj,Sen:2012dw}), which we now review, on how to subtract the zero mode contribution in the heat kernel.

Let $n_\CO^0$ be the number of zero modes of the operator $\CO$. We can write,
\bea
-\frac{1}{2} \: \ln \text{det}' \CO = \frac{1}{2} \int_{\epsilon}^{\infty} \: \frac{d \tau}{\tau} \: \big{(} \text{Tr} K (\tau) - n_\CO^0 \big{)}
\eea
where $\epsilon$ is a UV cutoff. At small $\tau$, we can employ the  Seeley-De Witt expansion for the heat kernel  which leads to
\bea
\text{Tr} K (\tau) = \frac{1}{(4 \pi)^{d/2}} \: \sum_{n = 0}^{\infty} \: \tau^{n - d/2} \: \int d^d x \: \sqrt{g} \: a_n (x,x).
\eea
Since non-zero eigenvalues of a standard Laplace-like  operator $\CO$ scale as $L^{-2}$, it is natural to redefine $\bar{\tau} = \tau/L^2$. The expression for the determinant of the operator $\CO$ can be rewritten as
\bea
-\frac{1}{2} \: \ln \text{det}' \CO = \frac{1}{2} \int_{\epsilon/L^2}^{\infty} \: \frac{d \bar{\tau}}{\bar{\tau}} \: \bigg{(} \sum_{n = 0}^{\infty} \: \frac{1}{(4 \pi)^{d/2}} \: \bar{\tau}^{n - d/2} \: L^{2 n - d} \: \int d^d x \: \sqrt{g} \: a_n (x,x) - n_\CO^0 \bigg{)}.
\eea
From the above expression it is clear that the logarithmic contribution to $\ln \text{det}' A$ comes only from the term $n = d/2$,
\bea
-\frac{1}{2} \: \ln \text{det}' \CO =  \bigg{(} \frac{1}{(4 \pi)^{d/2}} \: \int d^d x \: \sqrt{g} \: a_{d/2} (x,x) - n_\CO^0 \bigg{)} \: \text{log} L + \ldots.
\eea

On very general grounds of diffeomorphic invariance, it can be argued that in  odd-dimensional spacetimes, the coefficient $a_{d/2}$ vanishes \cite{Vassilevich:2003xt}. Therefore, the only contribution to the heat kernel comes from the zero modes in the form  $n_\CO^0$ above. Applied to our case, the one-loop contribution due to 11d supergravity  comes from the analysis of zero modes.

The  one-loop partition function can then be written schematically as
\be
Z_{\text{1-loop}}[\beta,\ldots]=\sum_{D}(-1)^D(\frac{1}{2}\log \text{det}'D)+\Delta F_0,
\label{Eq:oneloop}
\ee
where $D$ stands for kinetic operators corresponding to various fluctuating fields and $(-1)^D=-1$ for bosons and $1$ for fermions.  The prime indicates removal of the zero modes, which are accounted for separately by
\be
\Delta F_0=\log \int [d\phi]|_{D\phi=0},
\label{zm}
\ee
where $\exp(-\int d^dx \sqrt{g}\phi D\phi)=1$. The structure of the logarithmic term is then given by
\begin{equation}\label{logterm}
\log Z[\beta, \dots]=\sum_{\{D\}}(-1)^D(\beta_D-1)n_D^0\log L+ \Delta F_{\mathrm{Ghost}}+\cdots,
\end{equation}
where the ghost contributions are treated separately, as in \cite{Bhattacharyya:2012ye,Liu:2017vbl}, and $\beta_D$ is due to the integration over zero modes, Eq.~(\ref{zm}), in the path integral, as studied in various cases of logarithmic contributions to the black hole entropy and the one-loop partition function \cite{Sen:2011ba, Banerjee:2010qc, Banerjee:2011jp, Bhattacharyya:2012ye}.

It is worth noting that the coefficient of the logarithmic in $L$ term is independent of the UV cutoff, $\epsilon$, and, therefore, independent of the UV details of the theory -- this fact attest to the robustness of the logarithmic corrections to the black hole entropy. Whenever a microscopic UV theory, which in our case are supersymmetric field theories,  presents us with a prediction for the logarithmic coefficient, we can test if our macroscopic long distance gravity theory generates the same contribution.  

These  properties have, in fact, been already exploited in the context of the logarithmic corrections to BMPV black holes in  \cite{Sen:2012cj} whose logarithmic contributions come from an effective   five-dimensional supergravity theory. Analogously, the authors of  \cite{Bhattacharyya:2012ye}  successfully matched the logarithmic term in the large $N$ expansion of the ABJM free energy on $S^3$ with a gravity computation  performed in 11d supergravity  which essentially reduced to the contribution of a two-form zero mode.  More recently,  a similar  approach applied to the magnetically charged asymptotically $AdS_4$ black holes dual to the topologically twisted ABJM theory lead to perfect agreement with the field theory  \cite{Liu:2017vbl} .

\subsection{Zero-mode contributions}

As explained in the previous section, the computation of the one-loop effective action  in odd-dimensional spacetimes reduces to a careful treatment of the zero modes of  the relevant operators. When integrating over zero modes  for a kinetic operatore $D$, there is a  factor of $L^{\pm \beta_D}$ for each zero mode. The total contribution to the partition function from the zero modes is
\bea
\label{Eq:beta}
L^{\pm \beta_D \: n_D^0}.
\eea
In what follows we will discuss the coefficients $\beta_D$ and $n_D^0$ closely  following arguments already present in the literature. In particular, we are going to be most concerned with the effective theory on $AdS_4$ where a complete control of the various zero modes has been achieved  \cite{Sen:2012cj,Camporesi:1994ga}.  The zero modes we deal with in asymptotically $AdS$ spacetimes originate in modes that would have been pure gauge were it not for the fact that the gauge parameters are not normalizable.  It is interesting to note that the mathematical literature has its idiosyncratic and completely equivalent approach by way of $L^2$ cohomology \cite{10.2307/2042193,Donnelly1981}. We are going to closely follow the presentation of   \cite{Sen:2012cj} .

Typically, zero modes are associated with certain asymptotic symmetries. For example, with gauge transformations that do not vanish at infinity. The key idea in determining $\beta_D$ above in equation (\ref{Eq:beta}) is to find the right variables of integration and to count the powers of $L$ that such integration measure contributes when one starts from fields that would naturally be present in the action.

For example, let $A_{\mu}$ be a vector field in $d$-dimensional spacetime  and $g^{\mu \nu}$ be the  background metric which we assume can be written as $L^2 \: g^{(0)}_{\mu \nu}$; where $L$ is the radius of curvature and $g^{(0)}_{\mu \nu}$ is independent of $L$. The path integral over $A_{\mu}$ is normalized such that
\bea
\int [D A_{\mu}] \: \text{exp} \: \bigg{[} - \int d^{d} x \: \sqrt{\text{det} g } \: g^{\mu \nu} \: A_{\mu} \: A_{\nu} \bigg{]} = 1,
\eea
i.e. 
\bea
\int [D A_{\mu}] \: \text{exp} \: \bigg{[} - L^{d-2} \: \int d^3 x \: \sqrt{\text{det} g^{(0)} } \: g^{(0) \: \mu \nu} \: A_{\mu} \: A_{\nu} \bigg{]} = 1,
\eea

Then, the correctly normalized integration measure will be
\bea
\prod_{x, (\mu)} \: D \big{(} L^{(d-2)/2} \: A_{\mu} (x) \big{)}
\eea
Gauge fields zero modes are associated with deformations produced by the  gauge transformation with non-normalizable parameters $\delta A_\mu \propto \partial_\mu \Lambda(x)$.  Therefore, when integrating over vector zero modes one has
\bea
\beta_{A} = \frac{d-2}{2}.
\eea
Similarly we arrive to analogous expressions for various fields. For example, the expression for gravitons, gravitinos, and p-form fields has been discussed in \cite{Sen:2012cj}. Everything we need has been spelled out clearly in Sen's copious bibliography on the subject. In particular, we make heavy use of section 2 of  \cite{Sen:2012cj}.  Here, we will only need the expression for a 3-form potential as pertains to 11d supergravity.  To compute $\beta_{C_3}$ we assume similar scaling as before and obtain 
\be
\label{Eq:C3}
\beta_{C_3}=\frac{d-6}{2}.
\ee
For ease of visualization of the structure of the one-loop effective action, it is helpful to consider the  dimensional reduction from the 11d supergravity fields to $AdS_4$; we emphasize that the actual computation takes place in 11d and this is just a convenient book-keeping device. For the metric fluctuations we essentially have $G_{MN}=\{h_{\mu\nu}, h_{\mu n }, h_{mn}\}$, where the Greeek indices  $\mu, \nu$ are indices on $AdS_4$, the Latin indices  $m,n, $ denote directions in seven-dimensional manifold. The dimensional reduction of the metric leads to: one graviton in $AdS_4$, seven vector fields and a number of scalars. 

The other field of 11d supergravity is the 3-form potential $C_3$.  Recall that the general action for quantizing a $p$-form $A_p$ requires $p$ generalized ghosts \cite{Siegel:1980jj,Copeland:1984qk}. The gist of the argument, as succinctly explained in \cite{Siegel:1980jj}, is that when quantizing a $p$-form, $A_p$, one attempts to fix the invariance $A_p=d\Lambda_{p-1}$. However, the ghost arising by fixing a gauge,  acquires a gauge transformation since it is itself invariant under $\Lambda_{p-1}$ that are themselves exact. The prescription is cleverly summarized as -- ghosts themselves have ghosts \cite{Siegel:1980jj}.

The combined action for the $p$-form and its ghosts is given by \cite{Copeland:1984qk}:
\be 
S=-\frac{1}{2}\sum\limits_{j=0}^p\frac{1}{(p-j)!}(A_{p-j}, (\Delta_{p-j})^{j+1}A_{p-j}),
\ee
where the standard scalar product of forms is denoted by $(\cdot, \cdot)$ and $\Delta_{p-j}$ is the Hodge Laplacian. The $(p-j)$-forms $A_{p-j}$ is treated as a commuting field if $j$ is even and as an anticommuting field if $j$ is odd. The contribution to the one-loop effective action is thus
\be
\Gamma^{(1)}_p=-\frac{1}{2}\sum\limits_{j=0}^p(-1)^j(j+1)\ln \det \Delta'_{p-j},
\ee
where prime indicates removing of the zero modes.

Since we are computing the one-loop effective action in an odd-dimensional spacetime we know that the contribution to the logarithmic term can only come from the zero modes. Recall that in $AdS_{2M}$ there is only a $M$-form zero mode \cite{Camporesi:1995fb}; we are thus interested in tracking the contribution of the $2$-form zero mode present in asymptotically $AdS_4$ backgrounds.

The zero mode contribution, in turn, can only come from 2-form $A_{p-j=2}$ which corresponds for  $p=3$ to $ j=1$ and leads to the following one-loop contribution 
\bea
\label{Eq:Contribution2}
(-1)^j (\beta_{2}-1-1)n^0_{2} \ln L = (2-\beta_{2})n^0_2\ln L.
\eea

Given the backgrounds, the $3$-form potential of 11d supergravity can be decomposed as $C_{MNP}=\{C_{\mu\nu\rho}, C_{\mu\nu p}, C_{\mu n p}, C_{mnp}\}$, where Greek indices are legs on the asymptotic $AdS_4$ space and Latin indices are legs on $Y_7$.  Then a 2-form zero mode on the $AdS_4$ part can contribute if there is a $1$-form zero mode on $Y_7$. We will assume for now that such a one-form zero mode does not exist, that is, $b_1(Y_7)=0$ and proceed; this limitation is in accordance with the field theory conditions we have encountered. We will return to the slightly more general case toward the end of the section.

One might wonder if there are contributions arising form the quantization of the graviton. This problems has been explicitly addressed in, for example, \cite{Christensen:1979iy} and, given the gauge  invariance, requires the introduction of ghosts fields. In particular, there is a vector ghost, see equation (3.10) in  \cite{Christensen:1979iy}. However, the form of the operator in this case is $V^*_M(-g^{MN}\Box -R^{MN})V_N$ which does not admit zero modes due to background Einstein space we discuss: $R_{MN}\propto G_{MN}$. 
\subsection{The one-loop effective action and  logarithmic correction}

The most important ingredient in formulating the answer for the one-loop effective action is thus the number of two-form zero modes. Although our background is intrinsically eleven-dimensional, we can exploit the four dimensional point of view described in section \ref{Sec:holography}. Let us consider, for example the black holes that the action (\ref{Eq:4dGrav}) admits.

As can be see from equation (\ref{Eq:BlackHole}), the black hole we are interested in is an extremal one. It is known that for matters of thermodynamics, it is best to approach the computation of the effective action of the extremal solution through the computation in the non-extremal branch and then taking the limit to extremality. This prescription  has been discussed in detail in the context of the quantum entropy  function \cite{Sen:2008vm}   and used more recently  in a context similar to the one we consider here \cite{Liu:2017vbl}. The generic form of the non-extremal magnetically charged asymptotically  $AdS_4$ black hole with arbitrary genus $g$ horizon topology takes the form 
\be
ds^2=-f(r)dt^2 + \frac{dr^2}{g(r)}+h(r)ds^2(\Sigma_g), 
\ee
In principle the functions $f(r), g(r)$ and $h(r)$ will depend on the charges. 

Let us denote the number of 2-form zero modes of these solution by $n_2^0$. As explained in \cite{Liu:2017vbl}, $n_2^0$ is   the result of a regularized object and can be best understood as the properly defined in $L^2$ Euler characteristic. An interesting  application of such regularized Euler characteristic was explicitly presented in \cite{Larsen:2015aia} to elucidate aspects of quantum inequivalence in ${\cal N}=8$ gauged supergravity in four dimensions. The number of 2-form zero is 
\be
n_2^0=2(1-g).
\ee
It is important that this value is independent of the particular charges of the black hole. Therefore, as long as we approach the extremal solution through this branch we obtain the same result as for the non-extremal solution. Similarly, our computation applies for the one-loop quantum effective action of the Taub-Bolt-AdS${}_4$ solution discussed in section \ref{Sec:holography} as it admits the same embedding in eleven dimensional supergravity and has the same number of 2-form zero modes.

The full contribution to the logarithmic terms of the one-loop effective action is thus given only by the  2-form zero modes and we have: 
\be
\boxed{
\log Z_{1-loop}=(2-\beta_2)n_2^0\log L= (2-7/2)2(1-g)\log L = -(1-g)\log N},
\ee
where  according to the structure of the M5 brane solution we have $L^3\sim N$ (see Table \ref{AdS4/CFT3 from M2/M5-branes}). This result perfectly matches the field theory expectation and constitutes one of the main results of the manuscript.

Let us further discuss this result and understand its potential generalizations. In articular, we need to be aware of potential contributions coming from the fact that we are truly working in an eleven dimensional setup.  

Given that the only zero mode in $AdS_4$ is a 2-form and assuming that the solution is roughly of the form of warped products of  $AdS_4 \times M_3\times \tilde{S}^4$ we need to decompose the kinetic operator along these three subspaces. For the 2-form zero mode of $AdS_4$ to survive we need to have the corresponding part of the kinetic Laplace-like operator  also vanishing. The number of zero modes depends on the topology of the full space.


Let us now address the crucial role of the compactness of $M_3$. Given that $M_3$ is locally $\mathbb{H}^3$, one might assume naively that $M_3=\mathbb{H}_3$. This would imply that the 2-form zero mode in $AdS_4$ is lifted because there are no zero modes on $\mathbb{H}^3$.  Given that $\mathbb{H}^3$ is simply connected the De Rham intuition indicates that there might be a zero mode. However,  for a non-compact space, and in the context of $L^2$ cohomology, a constant function is not $L^2$-normalizable  and does not contribute\footnote{We thank Wenli Zhao for various clarifications on $L^2$ cohomology.}.  This would imply that there are no zero modes in the full solution and, therefore, no contribution to the logarithmic term.  This gravity intuition might inform attempts to wrap M5 branes on non-compact hyperbolic spaces, we do not pursue this direction in this manuscript. 


Let us return to $M_3$ compact and admitting a one-form zero mode.   We have assumed that $M_3$ is compact and  connected, that is, $b_0(M_3)=1$; similarly we have assume that $\tilde{S}^4$ is topologically a 4-sphere and therefore $b_0(\tilde{S}^4)=1$ and $b_4(\tilde{S^4})=1$ with all other Betti numbers for $\tilde{S}^4$ vanishing. 
Depending on the topology of $M_3$ there could also be other contributions to the coefficient of the logarithmic in $N$ term. For example, if $M_3$ admits one-form zero modes we could construct a 3-form zero mode which is the wedge product of the  2-form zero mode on $AdS_4$ and the one-form on $M_3$. This will contribute through the $C_3$ integration. Let us explicitly compute such contribution. Recall that the expression for $\beta_{C_3}$ given in equation (\ref{Eq:C3}) leads to $\beta_{C_3}=5/2$ in $d=11$ dimensions.  The contribution to the one-loop effective action following from the master equation (\ref{Eq:oneloop}) is: 

\bea
\log Z\big|_{C_3} &=& (-1)^1(\beta_{C_3}-1) n_{C_3}^{(0)}\log L \nonumber \\
&=& - \left(\frac{5}{2}-1\right) 2 (1-g) b_1 \log L\nonumber \\
&=& 3 (g-1) b_1 \log L\nonumber \\
&=& (g-1) b_1 \log N,
\eea
where in the last equality we have again translated from $L^3 \sim N$ according to Table \ref{AdS4/CFT3 from M2/M5-branes}. 

For completeness we note that the one-form zero mode on $M_3$ can not contribute through the one-form ghost determinants because there are no normalizable $0$-form in the asymptotically $AdS_4$ region.  Similarly the 3-form zero mode on $M_3$ can not contribute through the  $C_3$ integration.

The most general expression that we have is, therefore:

\be
\boxed{
\log Z_{1-loop}=(g-1)(1+b_1)\log N}.
\ee

It would be interesting to relax the $b_1(M_3)=0$ condition on the field theory side and compare with this gravity prediction for the logarithmic in $N$ term. Alternatively, this expression can be used as an IR consistency check for would-be UV expressions.

\section{Conclusions}\label{Sec:Conclusions}

In this manuscript we have considered partition functions of 3d field theories  denoted by ${\cal T}_N[M_3]$ which are obtained as the low energy limit of $N$ M5 branes wrapping a hyperbolic 3-manifold, $M_3$.   By exploiting the connection of the ${\cal T}_N[M_3]$  theory with  $PSL(N,\mathbb{C})$ Chern-Simons theory on $M_3$ we were able to produce expressions for the partition functions in the large $N$ limit including perturbative corrections to all orders in $1/N$.  This is an important achievement, especially in comparison with the state of the art of generic computations of the topologically twisted indices of other 3d field theories arising as the worldvolume theories of D2 or M2 branes. In those cases the field theory supersymmetric observables are only obtained at leading order in $N$ \cite{Hosseini:2016tor,Hosseini:2016ume}  or, with some numerical effort, at sub-leading order \cite{Liu:2017vll,Liu:2018bac}.

 One important sub-leading result obtained in the manuscript is a logarithmic in $N$ term on the field theory side. On the dual gravity side, the coefficient of the logarithmic in $N$ term is an IR window into the UV physics as eloquently stated by Sen \cite{Sen:2011ba}. In our case the UV physics of the gravity theory is provided by the field theory.  Exploiting the connection with Chern-Simons and results in the mathematical literature, we now have an analytic result for the coefficient of the $\log N$ contribution. This is a substantial improvement with respect to previous results in the literature of partition functions for generic ${\cal N}=2$ supersymmetric field theories in 3d. We have also computed the coefficient of the logarithmic in $N$ corrections using  exclusively the massless degrees of freedom of the dual  eleven dimensional supergravity describing the stack of M5 branes and found precise agreement with the field theory result.  Using these IR data we have a perfect match with the UV answer coming from the field theoretic analysis.  We have also demonstrated that the result is rather universal in the sense that it depends on a few topological aspects of the hyperbolic 3-manifold, $M_3$.  We have pointed out that improvements  on the field theory and gravity sides are possible. It would be interesting to better understand the field theory for arbitrary homology of the 3-manifold $M_3$. In particular, there is a gravity prediction for hyperbolic 3-manifolds of an arbitrary first Betti number $b_1$. The gravity side of the computation is easily extendable to more general cases and we expect more stringent tests to take place in the future. 

Given the nature of the field theory answer, it would be quite interesting to understand other terms in  the  $1/N$ expansion from the gravitational point of view. In particular, it would be quite interesting to provide a Wald entropy interpretation for various terms in the expansion of the topologically twisted index, see \cite{Hristov:2018lod,deWit:2018dix} for recent developments in understanding  the quantum entropy function from  AdS gravity side. 

Recall that in the context of the AdS/CFT correspondence the field theory provides the exact answer via the index to the gravity question of quantum entropy. Through AdS/CFT this amounts to having the UV complete answer to the question of microstates counting on the gravity side.  Understanding the structure of the indices in 3d supersymmetric field theory more broadly thus corresponds to uncovering the precise structure of the underlying string theory. Let us elaborate on this possibility of high precision holography  where the field theory is providing the analog of the full string theory partition function as was the case in \cite{Strominger:1996sh}. One ultimate goal of the program we pursue here is to achieve a full understanding of the asymptotic form of the partition function; similar to certain dyonic states in string theory   \cite{Dabholkar:2004yr,Sen:2007qy} where it was demonstrated that the quantum corrected macroscopic entropy agrees precisely with the microscopic counting for an infinite tower of fundamental
string states to all orders in an asymptotic expansion. In our  case we were aided by the relation to Chern-Simons theory for which there are many results in the mathematical literature which we can re-direct to our purpose.  It would be interesting to pursue this program for more general ${\cal N}=2$ superconformal field theories. 

The non-trivial issue of integrality which was crucial in previous approaches has been addressed here with explicit examples.  We hope to understand this aspect in a more general and formal manner, although the evidence for it is convincing enough. The number of states (quantum entropy) $\log d(Q_i, P_i)$  should be related to an integer number of states.  In the derivation of the 3d-3d relation in Table~\ref{3d/3d correspondence for twisted index}, we use a field theoretic construction of the 3d field theory $\mathcal{T}_{N}[M_3]$. It would be interesting to derive the relation directly from the 6d definition of the $\mathcal{T}_{N}[M_3]$ as done in \cite{Yagi2013,Lee2013,Cordova2017} for other supersymmetric partition functions. We leave some of these questions for future research.

\section*{Acknowledgments}
We are grateful to  J. Bae,  F. Benini, J. Hong, S. M. Hosseini, S. Kim, K. Lee,  J. T. Liu, N. Macpherson,  V. Rathee, V. Reys, A. Sen, 
M. Yamazaki, and W. Zhao for illuminating comments.   DG was partly supported by Samsung Science and Technology Foundation under Project Number SSTBA140208. NK is partially supported by the National Research Foundation of Korea (NRF) grant 2018R1D1A1B07045414.  LAPZ is partially supported by the US Department of Energy under Grant No. de-sc0007859.

\newpage
\appendix

\section{Analytic torsions on $M_3=(S^3 \backslash \mathbf{4}_1)_{P\mu + Q \lambda}$} \label{app : torsion computation}
In this appendix we  give an explicit expression for the analytic torsion $T^{\alpha}_{M_3} (\tau_{2n+1},N=2)$ for 3-maniofolds $M_3 = (S^3 \backslash \mathbf{4}_1)_{P \mu+Q\lambda}$ obtained from Dehn surgeries along a  `figure-eight' knot ($\mathbf{4}_1$). Refer to, for example, to \cite{Dimofte:2012qj,2015arXiv151100400P,2016arXiv160407490G} for recent mathematical developments on the topic.
\begin{figure}[htbp]
	\begin{center}
		\includegraphics[width=.2\textwidth]{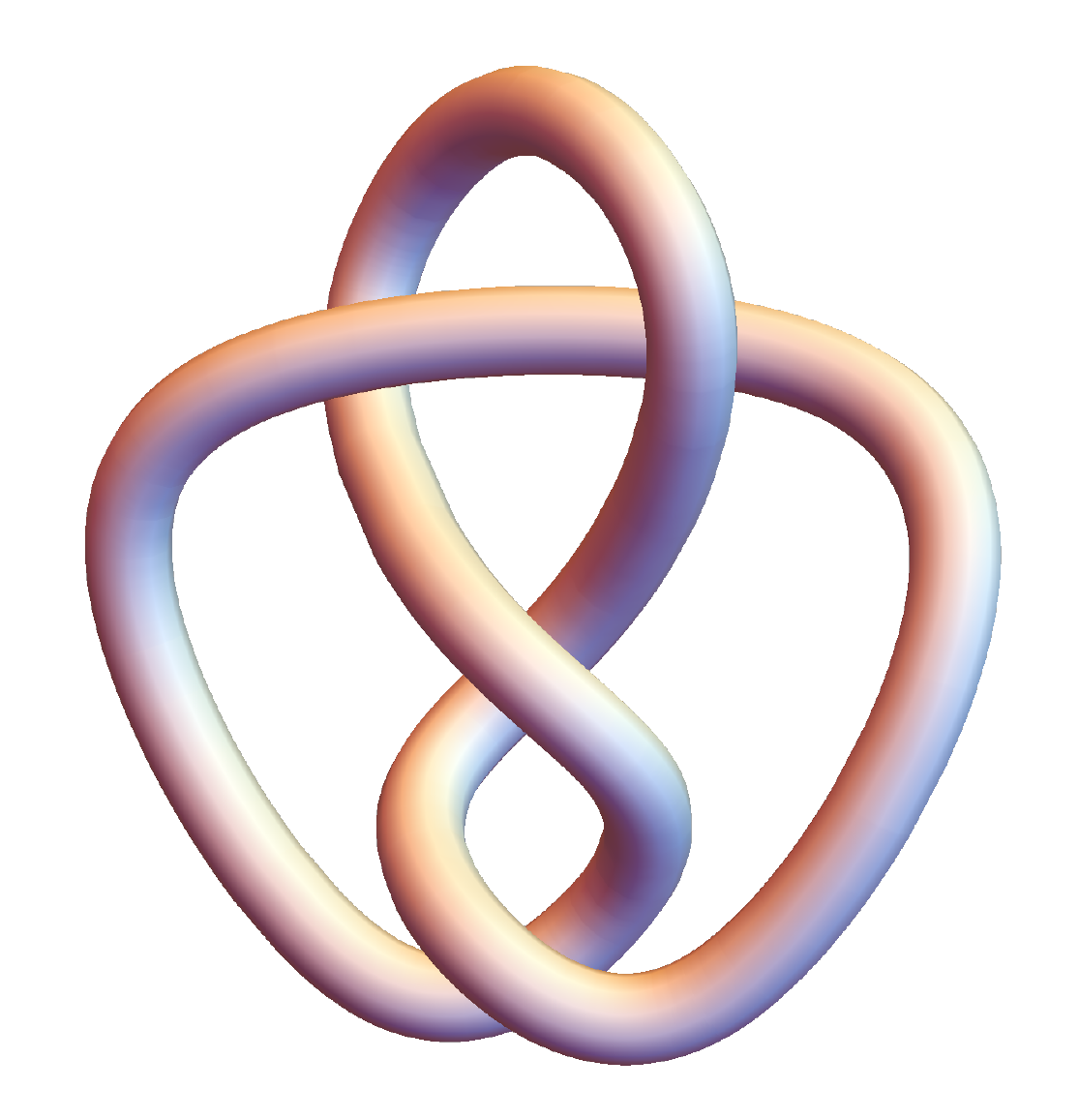}
	\end{center}
	\caption{Figure-eight knot, $K = \mathbf{4}_1$ }
	\label{fig:figure 8 knot}
\end{figure}
\paragraph{Closed 3-manifold from surgery along a knot $K$: }  One systematic way of constructing  closed 3-manifolds is using Dehn surgery along a knot $K$ in  3-sphere $S^3$. The Dehn surgery  can be done in two steps, drilling and  filling. First, we remove a tubular neighborhood of the knot $K$ and create a 3-manifold $S^3\backslash K$ called knot complement:
%
\begin{align}
\textrm{Drilling : }S^3 \backslash K := S^3 - (\textrm{Tubular neighborhood of a  knot $K$})\;.
\end{align}
The  3-manifold has a single torus boundary, which corresponds to the boundary of  removed tubular neighborhood of the knot. 
\begin{align}
H_1 \left(\partial (S^3\backslash \mathbf{4}_1),\mathbb{Z}\right) = H_1 (\mathbb{T}^2,\mathbb{Z}) = \mathbb{Z} \times \mathbb{Z} = \langle \mu, \lambda \rangle 
\end{align}
There is a canonical basis of boundary 1-cycles called meridian ($\mu$) and longitude ($\lambda$). $\mu$ is the generator of $H_1 (S^3\backslash K, \mathbb{Z})$ while $\lambda$ is a trivial element in the homology.
\begin{align}
\begin{split}
&H_1 (S^3\backslash K, \mathbb{Z}) = \mathbb{Z} = \langle \mu \rangle\;,
\\
& \lambda \textrm{ is a trivial element in $H_1 (S^3\backslash K, \mathbb{Z}) $}\;.
\end{split}
\end{align}
As the last step, we glue back to the removed solid-torus in a way that the boundary cycle $p\mu + q\lambda$ is glued to the shrinking cycle of the solid-torus. The procedure is called Dehn filling and the resulting closed 3-manifold will be denoted as
$ (S^3\backslash K )_{P \mu + Q \lambda} $:
\begin{align}
\begin{split}
\textrm{Dehn filling : } &(S^3\backslash K)_{P \mu+ Q\lambda} = \big{(} (S^3\backslash K) \cup (\textrm{solid-torus}) \big{)}/\sim \;,
\\
& (P\mu + Q\lambda) \sim (\textrm{shrinkable boundary 1-cycle of solid-torus})\;.
\end{split}
\end{align}

\paragraph{Fundamental group $\pi_1 (M_3)$:} The fundamental group of the figure-eight knot complement $S^3\backslash \mathbf{4}_1$ is
\begin{align}
\pi_1 (S^3 \backslash \mathbf{4}_1)  = \langle a,b : a b^{-1} a^{-1} b a = b a b^{-1} a^{-1} b   \rangle\;.
\end{align}
Its peripheral subgroup is
\begin{align}
\pi_1 (\partial (S^3 \backslash \mathbf{4}_1)) = \mathbb{Z} \times \mathbb{Z} = \langle \mathbf{m} := a, \mathbf{l} :=  a b^{-1} a b a^{-2} b a b^{-1}a^{-1} \rangle \subset \pi_1 (S^3\backslash \mathbf{4}_1)\;.
\end{align}
There is an isomorphism between $\pi_1 \left(\partial (S^3 \backslash \mathbf{4}_1)\right)$ and  $H_1 \left(\partial (S^3 \backslash \mathbf{4}_1),\mathbb{Z}\right)$:
\begin{align}
\mathbf{m}^P \mathbf{l}^Q \quad \leftrightarrow \quad P \mu + Q \lambda\;.
\end{align}
The fundamental group of a closed 3-manifold $M_3=(S^3 \backslash \mathbf{4}_1)_{P\mu + 
	Q \lambda}$ ($P,Q$ are co-prime integers) is 
\begin{align}
\pi_1 (M_3) = \{ \mathbf{m}^P \mathbf{l}^Q =a^P (ab^{-1} a ba^{-2}b a b^{-1}a^{-1})^Q =1  \} \cap \pi_1 (S^3\backslash \mathbf{4}_1) \;.
\end{align}
The closed  manifold is always hyperbolic except for the following 10 choices of $(P,Q)$'s, which are called exceptional slopes
\begin{align}
(P,Q) = (0,1),(1,0),(\pm 1, 1),(\pm 2,1),(\pm 3,1),(\pm 4, 1)\;.
\end{align} 
First homology of $M_3 = (S^3\backslash K)_{P \mu+ Q\lambda}$ is
\begin{align}
H_1 (M_3, \mathbb{Z}) = \mathbb{Z}_{|P|}\;.
\end{align}
%

\paragraph{$\chi^{\rm irred}(N=2,M_3)$ from  solving gluing equations:} The figure-eight knot complement can be  triangulated using two ideal tetrahedra \cite{thurston1979geometry}. The gluing equations for the ideal triangulation are  followings \cite{Dimofte:2012qj}
\begin{align}
\begin{split}
\textrm{Gluing equation I :} \quad z_i  z_i' z_i'' =-1 \;, \; z_i^{-1} +z_i''-1=0\;, \; z_1^2 z_2^2 z_1'' z_2'' = 1\;, \label{gluing-equations-N=2}
\end{split}
\end{align}
 Solutions to the gluing equations give irreducible $PSL(2,\mathbb{C}) = PGL(2,\mathbb{C})=GL(2,\mathbb{C})/\mathbb{C}^*$ flat connections on the knot complement with the  following holonomy matrices
\begin{align}
\begin{split}
&\mathbb{M} := \mathbf{A} :=P \exp \left(- \oint_a \mathcal{A} \right) = \bigg{[} \left(
\begin{array}{cc}
\frac{z_2}{z_1}+\frac{1}{z_2'} & -\frac{z_2}{z_1} \\
-\frac{1}{z_1 z_2'}+\frac{1}{z_2'} & \frac{1}{z_1 z_2'} \\
\end{array}
\right) \bigg{]} \sim_{\rm conj}  \bigg{[} \left(
\begin{array}{cc}
1 & 0\\
* & m \\
\end{array}
\right) \bigg{]}  \;, 
\\
&\mathbf{B}:= P \exp \left(- \oint_b \mathcal{A} \right) = \bigg{[} \left(
\begin{array}{cc}
1 & 0 \\
1- \frac{1}{z_2} & \frac{z_1'}{z_2} \\
\end{array}
\right) \bigg{]}\; ,
\\
&\mathbb{L} := \mathbf{A} \cdot \mathbf{B}^{-1} \cdot \mathbf{A} \cdot \mathbf{B}\cdot \mathbf{A}^{-2}\cdot \mathbf{B}\cdot \mathbf{A}\cdot \mathbf{B}^{-1}\cdot \mathbf{A}^{-1} = P \exp \left(- \oint_{\mathbf{l}} \mathcal{A} \right) \sim_{\rm conj} \bigg{[}\left(
\begin{array}{cc}
\ell^{-1} & 0\\
* & \ell \\
\end{array}
\right) \bigg{]} \;,	
\end{split} \nonumber
\end{align}
where
\begin{align}
m =- z_1 z_2 z_1'' \;,  \quad  \ell = \frac{1}{z_1^2 z_1''} \;.
\end{align}
In the above, $[A]$ denotes the equivalence class of a $2\times 2$ matrix $A$ under the $\mathbb{C}^* = \mathbb{C}\backslash \{0\}$ action
\begin{align}
[A] = [t A] \quad \; \textrm{for $t \in \mathbb{C}^*$}\;.
\end{align}
Through the gluing equations in \eqref{gluing-equations-N=2}, the $m$ and $\ell$ are constrained by the following algebraic equation
\begin{align}
A^{\mathbf{4}_1}_{\rm poly}(m,\ell) = 2+ \ell + \frac{1}\ell - m^2 +m + \frac{1}m - \frac{1}{m^2}=0\;.
\end{align}
The polynomial is called A-polynomial of figure-eight knot.  To obtain flat connections  on $M_3= (S^3\backslash \mathbf{4}_1)_{ p \mu + q \lambda}$, we additionally  impose the following conditions after having imposed gluing conditions in  \eqref{gluing-equations-N=2}
\begin{align}
\begin{split}
&\textrm{Gluing equation II : \; }[\mathbb{M}^P \cdot \mathbb{L}^Q] =  [ \mathbb{I}]\;. \label{gluing-equations-N=2-2}
\end{split}
\end{align}
Let $(\hat{z}_i, \hat{z}'_i , \hat{z}''_i)_{\alpha}$ be solutions for the gluing equations in \eqref{gluing-equations-N=2}  and \eqref{gluing-equations-N=2-2}. The number of solutions is finite and the each solution  give 
\begin{align}
\begin{split}
&\rho_\alpha \in \textrm{Hom}[\pi_1 (M_3)\rightarrow PSL(2,\mathbb{C})]\;, \quad \textrm{where}
\\
& \rho_\alpha (a) = \mathbf{A}|_{(z_i,z_i',z_i'') = (\hat{z}_i, \hat{z}'_i , \hat{z}''_i)_{\alpha}}\;, \quad  \rho_\alpha (b) = \mathbf{B}|_{(z_i,z_i',z_i'') = (\hat{z}_i, \hat{z}'_i , \hat{z}''_i)_{\alpha}}\;.
\end{split}
\end{align}
Not all solutions give different irreducible flat-connections and we need to further quotient by conjugation 
\begin{align}
\begin{split}
&\chi^{\rm irred}(N=2,M_3) \subset  \textrm{Hom}[\pi_1 (M_3)\rightarrow PSL(2,\mathbb{C})]/(\textrm{conj})
\\
& = \{ \rho_\alpha :  (\hat{z}_i, \hat{z}'_i , \hat{z}''_i)_{\alpha} \textrm{ is a solution of gluing equations in \eqref{gluing-equations-N=2} and \eqref{gluing-equations-N=2-2}}  \big{\}}/(\textrm{conj}) \;.
\end{split}
\end{align}

\paragraph{ ${\bf Tor}^{(\alpha)}_{M_3} [\tau_{\rm adj},N=2]$  from state-integral model:} The analytic torsion ${\bf Tor}_{S^3\backslash  \mathbf{4}_1}[\tau_{\rm adj },N; P\mu + Q\lambda]$ depends on the choice of a primitive boundary 1-cycle, $P \mu+Q \lambda$ with co-primes $(P,Q)$. The torsion for $N=2$  can be computed as \cite{Dimofte:2012qj}
\begin{align}
\begin{split}
&{\bf Tor}_{S^3\backslash  \mathbf{4}_1}[\tau_{\rm adj },N=2; P\mu + Q\lambda] 
\\
&= \det \bigg{[} \left(
\begin{array}{cc}
2 & 2 \\
\frac{P}{2}-2 Q & \frac{P}{2} \\
\end{array}
\right) \left(
\begin{array}{cc}
z_1'' & 0 \\
0 & z_2'' \\
\end{array}
\right) +  \left(
\begin{array}{cc}
1 & 1 \\
\frac{P}{2}-Q & 0 \\
\end{array}
\right)   \left(
\begin{array}{cc}
1/z_1 & 0 \\
0 & 1/z_2 \\
\end{array}
\right) \bigg{]} z_1 z_2 \;. \label{41 torsion from state-integral model}
\end{split}
\end{align}
Using the formula in \eqref{Torsion under Dehn filling}, the torsion ${\bf Tor}^{(\alpha)}_{M_3} [\tau_{\rm adj},N=2]$ for a flat-connection $\rho_\alpha\in \chi^{\rm irred}(N=2,M_3)$ on  the Dehn filled closed 3-manifold $M_3= (S^3\backslash \mathbf{4}_1)_{P \mu +Q \lambda}$ is given by 
\begin{align}
{\bf Tor}^{(\alpha)}_{M_3 } [\tau_{\rm adj},N=2] = \frac{{\bf Tor}_{S^3\backslash  \mathbf{4}_1}^{(\alpha)}[\tau_{\rm adj },N=2; P\mu + Q\lambda]}{\big{(}1-(m_{\alpha})^{R} (l_{\alpha})^{2S} \big{)}\big{(}1-(m_{\alpha})^{-R} (l_{\alpha})^{-2S} \big{)}}\;. \label{41 torsion under Dehn filling}
\end{align}
Here integers $(R,S)$ are chosen such that
\begin{align}
\left(
\begin{array}{cc}
P & Q \\
R & S \\
\end{array}
\right) \in SL(2,\mathbb{Z})\;.
\end{align}
The choice is not unique but can be shifted as follows: 
\begin{align}
(R,S ) \rightarrow (R,S)  + \mathbb{Z} (P,Q)\;.
\end{align}
Note that the torsion in \eqref{41 torsion under Dehn filling} is invariant under the shift due to the gluing equations in \eqref{gluing-equations-N=2-2}. 
\paragraph{Example : $M_3=(S^3\backslash \mathbf{4}_1)_{5\mu + \lambda}$} The are 4 $PSL(2,\mathbb{C})$ flat-connections, $\rho_{\alpha=1,\ldots,4}$, in $\chi^{\rm irred}(N=2,M_3)$. Giving numerical expressions of the flat connections
\begin{align}
\begin{split}
\alpha =1\;, \quad &(z_1, z_2)=(0.169304\, +2.39229 i,0.80957\, +0.0692817 i)\;,
\\
&\rho (a) = \left(
\begin{array}{cc}
1.61306\, -0.169296 i & -1.02826+0.548774 i \\
0.409747\, +0.61154 i & 0.175049\, -0.232062 i \\
\end{array}
\right)\;,
\\
&\rho (b) = \left(
\begin{array}{cc}
1.20331\, -0.780836 i & 0. \\
-0.1903+0.302934 i & 0.584796\, +0.379478 i \\
\end{array}
\right)\;,
\\
\alpha =2\;, \quad & (z_1, z_2)=(0.169304\, -2.39229 i,0.80957\, -0.0692817 i)\;,
\\
&\rho (a) = \left(
\begin{array}{cc}
1.61306\, +0.169296 i & -1.02826-0.548774 i \\
0.409747\, -0.61154 i & 0.175049\, +0.232062 i \\
\end{array}
\right)\;,
\\
&\rho (b) =\left(
\begin{array}{cc}
1.20331\, +0.780836 i & 0. \\
-0.1903-0.302934 i & 0.584796\, -0.379478 i \\
\end{array}
\right)\;,
\\
\alpha =3\;, \quad & (z_1, z_2) = (-0.544322-0.324476 i,-0.544322+0.324476 i)\;,
\\
&\rho (a) =\left(
\begin{array}{cc}
-0.245108-1.46992 i & 0.409586\, +0.48354 i \\
-0.409586-2.4563 i & 0.574064\, +1.46992 i \\
\end{array}
\right)\;,
\\
&\rho (b) =\left(
\begin{array}{cc}
0.164478\, +0.986381 i & 0. \\
-0.409586+2.4563 i & 0.164478\, -0.986381 i \\
\end{array}
\right)\;,
\\
\alpha =4\;, \quad & (z_1, z_2) = (0.0654485\, +0.807157 i,0.0654485\, -0.807157 i)\;,
\\
&\rho (a)=\left(
\begin{array}{cc}
0.237556\, -0.468055 i & -0.690139-0.423667 i \\
0.690139\, -1.35978 i & -1.14272+0.468055 i \\
\end{array}
\right)\;,
\\
&\rho (b) =\left(
\begin{array}{cc}
-0.452583+0.891722 i & 0. \\
0.690139\, +1.35978 i & -0.452583-0.891722 i \\
\end{array}
\right)\;.
\end{split} 
\label{4 flat connections}
\end{align}
The $\alpha=1$ and $\alpha=2$ corresponds to $\alpha = (\overline{\rm geom})$ and $\alpha = (\rm geom)$  respectively. The analytic torsion for these flat-connections can be computed using equations in \eqref{41 torsion from state-integral model} and \eqref{41 torsion under Dehn filling} :
\begin{align}
\begin{split}
&{\bf Tor}^{(\alpha)}_{M_3}[\tau_{\rm adj},N=2]
\\
&= \{ -1.905381-0.568995 i,\; -1.905381+0.568995 i,\; 2.570846,\; 1.739916 \} \label{Torsion-thurston-N=2}
\end{split}
\end{align}
\paragraph{${\bf Tor}_{M_3}[\tau_{2n+1},N=2]$ from Fox calculus:} According to the Cheeger-Muller theorem, the analytic Ray-singer torsion  is actually equivalent to the Reidemeister torsion. The Reidemeister torsion is a purely combinatorial invariant and the quantity on knot complement can be computed from  Fox differential calculus on its fundamental group.
For example, the torsion ${\bf Tor}^{(\alpha)}_{S^3 \backslash \mathbf{4}_1}[\tau_{2n+1},N;\lambda]$ can be given as \cite{2016arXiv160407490G}.
\begin{align}
\begin{split}
&\Delta(\tau_{2n+1};\rho_\alpha) =\frac{ \det \left(I_{2n+1} -t^{-1} A_n B_n^{-1}A_n^{-1}+A_n B_n^{-1} A_n^{-1} B_n - t B_n +B_n A_nB_n^{-1}A_n^{-1}\right)}{\det (t I_{2n+1} - B_n)}\;,
\\
&A_{n} := \tau_{2n+1} (\rho_\alpha (a))\;, \quad  B_{n} := \tau_{2n+1} (\rho_\alpha (b))\;, 
\\
&{\bf Tor}^{(\alpha)}_{S^3 \backslash \mathbf{4}_1}[\tau_{2n+1}, N=2;\lambda] = \lim_{t\rightarrow 1} \frac{\Delta (\tau_{2n+1}, t;\rho_\alpha)}{t-1}. \label{torsion from fox}
\end{split}
\end{align}
Then, the torsion with respect to the general primitive boundary 1-cycle, $P \mu+Q \lambda$, is given by the  following transformation rule \cite{2015arXiv151100400P}
\begin{align}
\begin{split}
&{\bf Tor}^{(\alpha)}_{S^3 \backslash \mathbf{4}_1}[\tau_{2n+1}, N=2;P\mu + Q\lambda]  \;,
\\
& = \frac{\partial (\frac{P}2  \log m_\alpha + Q \log \ell_\alpha)}{\partial \log \ell_\alpha} {\bf Tor}^{(\alpha)}_{S^3 \backslash \mathbf{4}_1}[\tau_{2n+1}, N=2;\lambda]  \;,
\\
&= \left(- \frac{P}2\frac{ \ell \partial_{\ell} A^{\mathbf{4}_1}_{\rm poly}(m,\ell)}{ m \partial_m A^{\mathbf{4}_1}_{\rm poly}(m,\ell)}\bigg{|}_{m = m_\alpha, \ell = \ell_\alpha}+Q \right){\bf Tor}^{(\alpha)}_{S^3 \backslash \mathbf{4}_1}[\tau_{2n+1}, N=2;\lambda]\;,
\\
&=\left( P \frac{(\ell_\alpha - \frac{1}\ell_\alpha) m_\alpha^2}{(m_\alpha^2-1)(4-2m_\alpha+4m_\alpha^2)}+Q\right) {\bf Tor}^{(\alpha)}_{S^3 \backslash \mathbf{4}_1}[\tau_{2n+1}, N=2;\lambda]\;. \label{torsion on 41 for general reps}
\end{split}
\end{align}
Using the transformation rule of torsion under the Dehn filling \cite{2015arXiv151100400P}, we finally have
\begin{align}
{\bf Tor}^{(\alpha)}_{M_3=(S^3\backslash \mathbf{4}_1)_{P\mu + Q \lambda}} [\tau_{2n+1},N=2] = \frac{{\bf Tor}_{S^3\backslash  \mathbf{4}_1}^{(\alpha)}[\tau_{ 2n+1},N=2; P\mu + Q\lambda]}{\prod_{a=1}^n \big{(}1-(m_{\alpha})^{a R} (l_{\alpha})^{2 a S} \big{)}\big{(}1-(m_{\alpha})^{-a R} (l_{\alpha})^{-2a S} \big{)}}\;. \label{torsion on Dehn filled 41 for general reps}
\end{align}
\paragraph{Example : $M_3 = (S^3\backslash \mathbf{4}_1)_{5\mu+\lambda}$} Using the above formule in  \eqref{4 flat connections},\eqref{torsion from fox},\eqref{torsion on 41 for general reps} and \eqref{torsion on Dehn filled 41 for general reps}, we can compute the torsions and their numerical values are
\begin{align}
\begin{split}
&\{\log |{\bf Tor}_{M_3}^{\rm geom}[\tau_{2n+1}, N=2]|\}_{n=1}^\infty
\\
&=\{0.6873,1.5033,3.3932,5.8423,8.9316,12.777,17.120,
\\
&\qquad 22.108,27.740,33.983,40.856,48.354,56.475,65.222,\ldots\}
\end{split}
\end{align}
This series shows the expected asymptotic behavior in \eqref{N=2 torsion asymptotic}
\begin{align}
\begin{split}
&\big{\{}\log |{\bf Tor}_{M_3}^{\rm geom}[\tau_{2n+1}, N=2]|- \frac{(n^2+n)}\pi \textrm{vol}(M_3)\big{\}}_{n=1}^\infty
\\
&=\big{\{} 0.0626,-0.3708,-0.3552,-0.4052,-0.4397,-0.3424,-0.3732,
\\
&\quad  -0.3826,-0.3741,-0.3784,-0.3780,-0.3762,-0.3772,-0.3773,\ldots\big{\}}
\end{split}
\end{align}
The hyperbolic volume of the 3-manifold is \cite{SnapPy}
\begin{align}
\textrm{vol}(M_3) = 0.98136882889\ldots\;.
\end{align}
According to \eqref{N=2 torsion asymptotic}, the above series is expected to approach to a constant given by 
\begin{align}
\begin{split}
&\log |{\bf Tor}_{M_3}(N=1)| -\sum_{[\gamma]}\sum_{k=1}^{\infty} \log |1- e^{-k \ell_{\mathbb{C}}(\gamma) }|
\\
& = |{\bf Tor}_{M_3}^{\rm geom}[\tau_{5}, N=2]| - \frac{6}\pi \textrm{vol}(M_3)-\sum_{[\gamma]}\sum_{k=3}^{\infty} \log |1- e^{-k \ell_{\mathbb{C}}(\gamma) }|
\\
&= -0.3708 -\sum_{[\gamma]}\sum_{k=3}^{\infty} \log |1- e^{-k \ell_{\mathbb{C}}(\gamma) }|\;.
\end{split}
\end{align}


\bibliographystyle{JHEP}
\bibliography{ref}


\end{document}